\documentclass{article}

\usepackage{graphicx}
\usepackage{rotating}
\usepackage{amsmath,amssymb}

\title{Superconducting Lanthanum Nickel Oxides}
\author{
  Hiroya Sakurai and Yoshihiko Takano\\
  National Institute for Materials Science 
}

\date{February 19, 2026}

\begin{document}

\maketitle



\section*{Abstract}
In 2023, superconductivity in La$_3$Ni$_2$O$_7$ was discovered under high pressures above approximately 14~GPa. 
In addition to its high transition temperature ($T_{\mathrm{c}} \simeq 80$~K), the structural resemblance to high-$T_{\mathrm{c}}$ cuprates has strongly stimulated research, soon followed by the discovery of superconductivity in La$_4$Ni$_3$O$_{10}$. 
These compounds belong to the Ruddlesden--Popper phases, comprising double- and triple-layered NiO$_2$ square lattices separated by LaO rock-salt slabs.  

Research on these systems has rapidly developed along three major directions, as in other prominent families of superconductors such as the cuprates and iron arsenides: expanding the chemical variety of compounds, enhancing $T_{\mathrm{c}}$ through elemental substitution, and elucidating the superconducting mechanism. 
These challenges, being closely interconnected, continue to drive the field. 
The clarification of the pairing mechanism encounters a particular difficulty, since the key experiments must be performed under high pressures. 
This situation highlights the significance of developing nickel oxides that exhibit superconductivity at much lower pressures, ideally at ambient pressure, which would in turn broaden the scope of chemical tuning and detailed physical characterization.  

In this context, it is timely and meaningful to summarize the present state of knowledge. 
Here, we emphasize sample synthesis and characterization, which are already well established and often decisive for progress in unconventional superconductors, while providing a brief overview of the currently available electronic properties.  

\section{Introduction}

In 2023, La$_3$Ni$_2$O$_7$ was discovered to exhibit superconductivity under pressures above approximately 14 GPa \cite{SunNature2023}. 
The transition temperature reaches nearly 80 K. 
In addition to this relatively high $T_{\mathrm{c}}$, the compound shares a structural feature with high-$T_{\mathrm{c}}$ cuprates, as revealed shortly thereafter, which triggered intense research on this compound and related materials. 
Indeed, superconductivity was soon reported in La$_4$Ni$_3$O$_{10}$ above 33 GPa \cite{SakakibaraPRB2024}, and several new nickel oxides with similar structural features have since been synthesized to explore superconductivity, despite the experimental challenges under high pressure \cite{PuphalPRL2024,ChenJACS2024,WangInorgChem2024,LiPRM2024,YamaneACC2025}.

\begin{figure}
\includegraphics[width=8cm]{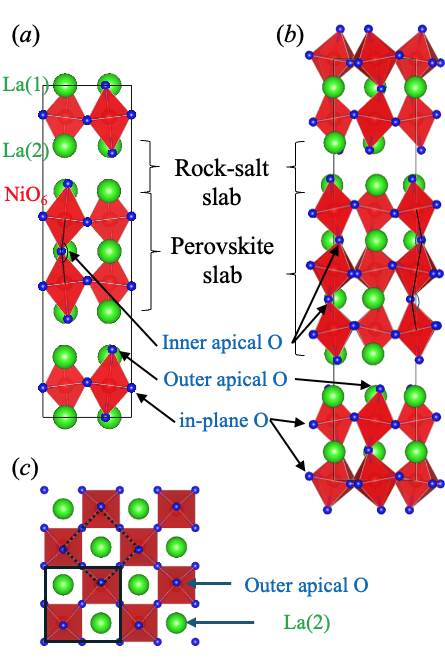}
\caption{\label{struct} 
Crystal structures of La$_3$Ni$_2$O$_7$ (a) and La$_4$Ni$_3$O$_{10}$ (b), with orthorhombic $Amam$ and monoclinic $P2_1/a$ symmetries, respectively. 
Panel (c) shows a top view of the perovskite and rock-salt slabs. 
Green, red, and blue circles represent La, Ni, and O atoms, respectively. 
Solid and dashed black lines in panel $c$ indicate the actual unit cell of La$_3$Ni$_2$O$_7$ and a virtual unit cell assuming tetragonal $I4/mmm$ symmetry, respectively.
}
\end{figure}

La$_3$Ni$_2$O$_7$ and La$_4$Ni$_3$O$_{10}$ crystallize in the Ruddlesden--Popper (RP) phase, with the general formula La$_{n+1}$Ni$_n$O$_{3n+1}$. 
Their crystal structure consists of alternating LaNiO$_3$ perovskite slabs and LaO rock-salt slabs, as shown in Fig.~\ref{struct}. 
In the perovskite slabs, NiO$_6$ octahedra are corner-sharing, forming $n$-layer NiO$_2$ square lattices, where Ni ions occupy the sites equivalent to Cu in the CuO$_2$ planes of cuprates. 
It should be noted that the Ni valence in the RP series increases from divalent to trivalent as $n$ increases. 
For example, La$_2$NiO$_4$ ($n=1$), which contains only divalent Ni ions, is a two-dimensional antiferromagnetic insulator with a charge-transfer energy gap \cite{AeppliPRL1988,IdoPRB1991}, similar to the parent compounds of cuprates. 
Upon partial substitution with trivalent Ni ions, this compound exhibits a charge and spin stripe state, which has been extensively studied for its possible relevance to stripe order in cuprates \cite{TranquadaNature1995}. 
On the other hand, LaNiO$_3$ ($n=\infty$), which contains only trivalent Ni ions, remains metallic down to the lowest temperatures, though it can become insulating when the bandwidth is reduced by substituting La with smaller lanthanide ions \cite{MedardeJPCM1997}. 
Furthermore, LaNiO$_3$ can be regarded as the parent compound of infinite-layer nickelate superconductors, which have also attracted much attention due to their electronic states analogous to those of cuprates \cite{LiNature2019,SakakibaraPRL2020}. 
The Ni valence in the infinite-layer nickelates lies between monovalent and divalent, in contrast to La$_3$Ni$_2$O$_7$ and La$_4$Ni$_3$O$_{10}$, which are situated on the opposite side of the phase diagram with respect to La$_2$NiO$_4$ \cite{TakegamiPRB2024}.  

These diverse electronic states indicate that the superconductivity in La$_3$Ni$_2$O$_7$ and La$_4$Ni$_3$O$_{10}$ is likely unconventional, rooted in their unique electronic structures. 
The purpose of this review is to summarize the experimental results on these nickel oxides available at present, with the hope of stimulating further research that will lead to the development of a wider variety of related materials and a deeper understanding of their rich physical properties.

\section{Crystal Structure}
\subsection{Ruddlesden--Popper phase}

The crystal structure of the Ruddlesden--Popper (RP) phase, A$_{n+1}$M$_{n}$X$_{3n+1}$, consists of AMX$_3$ perovskite slabs and AX rock-salt slabs, as described above for La$_{n+1}$Ni$_{n}$O$_{3n+1}$. 
To highlight this, the chemical composition can also be expressed as (AX)(AMX$_3$)$_n$. 
The ideal structure has tetragonal $I4/mmm$ symmetry for $n \neq \infty$. 
Within each slab, the A and X atoms adopt a cubic close-packed arrangement, whereas the interface between the slabs deviates from close packing, as shown in Fig.~\ref{struct}. 
In the perovskite slabs, the M ions form $n$ layers of square lattices as previously mentioned.

\begin{figure}
\includegraphics[width=8cm]{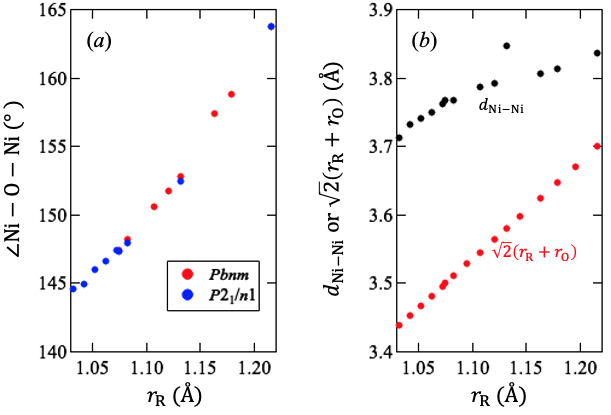}
\caption{\label{angle} 
(a) Average Ni--O--Ni bond angle in perovskite nickel oxides, and (b) average Ni--Ni distance together with $\sqrt{2}(r_R+r_O)$, plotted as functions of the ionic radius of the rare-earth ion for ninefold coordination estimated by Shannon \cite{ShannonActaCrystA1976}.
}
\end{figure}

The ideal perovskite AMX$_3$ ($n=\infty$) has cubic $Pm\bar{3}m$ symmetry, and its lattice constant corresponds to the distance between neighboring M atoms. 
For smaller A ions, the structure is stabilized by rotations of the AMX$_6$ octahedra. 
More specifically, when the tolerance factor $t = (r_A + r_X)/\sqrt{2}(r_M + r_X)$ (with $r_X$ the ionic radius of X) is larger than about 0.9, the ideal perovskite structure is favored. 
For smaller $t$, the MX$_6$ octahedra tilt to reduce the A-site volume \cite{GlazerActaCryst1972}. 
Typically, the LaAlO$_3$-type rhombohedral ($R\bar{3}c$) structure appears for relatively large $t$, while the GdFeO$_3$-type orthorhombic ($Pbnm$) structure occurs for smaller $t$, in the range $0.7 < t < 0.9$. 
In fact, LaNiO$_3$ adopts the LaAlO$_3$-type distortion, whereas $R$NiO$_3$ ($R$ = rare earths) perovskites with smaller ionic radii show the GdFeO$_3$-type structure for $R =$ Pr--Dy, and further distorted monoclinic ($P2_1/n$) structures for $R =$ Dy--Lu \cite{MedardeJPCM1997}. 
In these nickelate perovskites, the Ni--O bond length is almost independent of the rare-earth ion \cite{MedardeJPCM1997}, with NiO$_6$ octahedra behaving nearly as rigid units. 
Only the average Ni--O--Ni bond angle varies systematically with the ionic radius of $R$, as illustrated in Fig.~\ref{angle}$a$.

The rock-salt structure is realized for $r_A/r_X = 0.41$--0.73 \cite{BlossBook}, which corresponds to the condition that A atoms occupy the octahedral sites created by close-packed X atoms. 
However, many rock-salt compounds, such as CsF and BaO, exhibit significantly larger values than 0.73, indicating that for large A ions, both A and X atoms contribute comparably to the close packing. 
This applies to LaO rock-salt slabs, since $r_{\mathrm{La}}/r_{\mathrm{O}} = 0.87 > 0.73$. 
The lattice constant of an ideal rock-salt slab can be estimated as $a_{\mathrm{RS}} = 2(r_A + r_X)$, which matches the diagonal spacing of the square M lattice in the ideal RP structure. 
Thus, no mismatch arises between the rock-salt and perovskite slabs when $\sqrt{2}(r_A+r_X)$ equals the M--M distance. 
Because the X atoms in the rock-salt slabs are shared with the perovskite slabs, the way to tilt the MX$_6$ octahedra is severely restricted in RP phases with $n \neq \infty$. 
This likely explains why only a limited number of RP phases with $n=2$ or $3$ are known. 
For example, with A = La and X = O, only La$_3$Ni$_2$O$_7$, La$_4$Ni$_3$O$_{10}$, and La$_4$Co$_3$O$_{10}$ have been reported when containing only a single transition-metal element, which is surprising given that LaMO$_3$ perovskites exist for all $3d$ transition metals.
The variety of A ions for Ni oxides is also limited, likely becausue the mismatch between the perovskite and rock-salt slabs expands for smaller A ions as suggested in Fig. \ref{angle}$b$.
For a single A ion, no $n=2$ Ni oxide was known except for $R$ = La, and for $n=3$, only $R$ = Pr and Nd compounds were reported \cite{ZhangJSSC1995}, which is in sharp contrast with the variety of ANiO$_3$ perovskite \cite{MedardeJPCM1997}.
Unlike the case of perovskites, no simple and widely applicable parameter like the tolerance factor has been established for RP phases, presumably because it is difficult to describe the matching between perovskite and rock-salt slabs, although several attempts have been made \cite{SharmaBMS1998,GanguliJSSC1979,ShinAM2025}.

\subsection{La$_3$Ni$_2$O$_7$ and La$_4$Ni$_3$O$_{10}$}

For La$_3$Ni$_2$O$_7$, three types of crystal structures at ambient pressure and room temperature have been widely accepted: tetragonal $I4/mmm$ (\#139) \cite{ZhangJSSC1994n1}, orthorhombic $Fmmm$ (\#69) \cite{ZhangJSSC1994n1,SasakiJPSJ1997,AmowSSI2006RFR14,YuCEJ2016}, and orthorhombic $Amam$ (\#63) \cite{LingJSSC1999,VoroninNIMPRA2001,DeminaInorgMater2005,KiselevInorgMater2007}. 
Other symmetries have also been suggested \cite{KiselevChimicaTechnoActa2019,SongJMaterChemA2020}, but they are not generally accepted. 
The tetragonal symmetry is observed for oxygen-deficient La$_3$Ni$_2$O$_{6.35}$, suggesting that it is stabilized when the oxygen content is significantly below the stoichiometric value. 
By contrast, the origin of the two orthorhombic structures remains unclear, and in some samples both are observed to coexist \cite{XieSciBull2024}. 
Care should be taken in structure determination, as good refinement of X-ray diffraction (XRD) data with the $Fmmm$ model does not necessarily mean that the true structure is $Fmmm$ \cite{CarvalhoJMC1997}.

In the orthorhombic structures, the $a$ and $b$ axes correspond to the diagonal directions of the $c$-plane unit cell of the $I4/mmm$ structure, so their lengths are approximately $\sqrt{2}$ times the $a$ parameter of the tetragonal phase. 
In the $Fmmm$ structure, NiO$_6$ octahedra alternately contract and expand along the $a$ and $b$ axes without tilting from the $c$ axis, and the Ni--O--Ni bonds along $c$ remain linear. 
In contrast, in the $Amam$ structure, the Ni--O--Ni bonds are bent to about 168$^{\circ}$ due to tilting of the NiO$_6$ octahedra. 
It is widely accepted that the superconducting samples have the orthorhombic $Amam$ structure at ambient pressure, as superconductivity is thought to emerge when this bending is removed by a pressure-induced structural transition.

The orthorhombic $Amam$ phase of La$_3$Ni$_2$O$_7$ becomes unstable at around 10 GPa at room temperature, undergoing a transition to the orthorhombic $Fmmm$ structure near 15 GPa \cite{SunNature2023}.  
This structural transition has been supported by several XRD measurements \cite{SunNature2023,WangJACS2024,WangInorgChem2025} and DFT calculations \cite{SunNature2023}, although the slopes of the phase boundary reported in these studies are not consistent with each other \cite{WangJACS2024,WangInorgChem2025}.  
As mentioned above, distinguishing these structures by XRD patterns is difficult, particularly near the phase boundary, as the transition is first-order and the phases may coexist.
In general, DFT calculations do not account for effects arising at finite temperatures; thus, careful characterization is required to accurately determine the structural transition.  

At low temperatures under high pressure, the $Fmmm$ structure further transforms into the tetragonal $I4/mmm$ structure, accompanied by a clearer change in the XRD pattern \cite{WangJACS2024}, strongly suggesting that superconductivity occurs in the tetragonal phase, the ideal structure of the RP phase.  
Since the tetragonal $I4/mmm$ structure is also stabilized at high temperatures above approximately 450$^{\circ}$C under ambient pressure \cite{SongJMaterChemA2020}, the $Fmmm$ phase, if it exists, may be inserted between the $I4/mmm$ phases or separates them.  
For La$_2$PrNi$_2$O$_7$, the $Amam$ structure transforms directly into the $I4/mmm$ structure even around room temperature, without passing through the $Fmmm$ phase \cite{WangNature2024}.

Also for La$_4$Ni$_3$O$_{10}$, three types of crystal structures are recognized at room temperature and ambient pressure: orthorhombic $Fmmm$ (\#69) \cite{ZhangJSSC1995}, orthorhombic $Bmeb$ (\#64) \cite{LingJSSC1999,VoroninNIMPRA2001,DeminaInorgMater2005,KiselevInorgMater2007,ZhangPRM2020,YuanJCG2024}, and monoclinic $P2_1/a$ (\#14) \cite{NagellSSI2017,ZhangPRM2020,YuanJCG2024}.  
At higher temperatures above approximately 750$^{\circ}$C, the tetragonal $I4/mmm$ structure appears \cite{SongJMaterChemA2020,NagellSSI2017}.  
The orthorhombic $Bmeb$ structure can be metastable \cite{ZhangPRM2020}; 
in fact, it is reported to occur at temperatures between approximately 600$^{\circ}$C and 700$^{\circ}$C \cite{NagellSSI2017}.  
In the $Fmmm$ structure, the bridging angle of Ni--O--Ni along the $c$ axis is straight, as in La$_3$Ni$_2$O$_7$, whereas in the orthorhombic $Bmeb$ and monoclinic $P2_1/a$ structures, the angle bends from 180$^{\circ}$.  
After the discovery of superconductivity in La$_4$Ni$_3$O$_{10}$ \cite{SakakibaraPRB2024}, it was found that the superconductivity also occurs in the $I4/mmm$ structure \cite{ZhuNature2024,ZhangPRX2025,LiSCPMA2024}.  
Differently from La$_3$Ni$_2$O$_7$, this ideal RP structure is preserved even at room temperature under pressures above approximately 33 GPa \cite{ZhuNature2024,ZhangPRX2025,LiSCPMA2024}.  

The occurrence of the $I4/mmm$ structures in La$_3$Ni$_2$O$_7$ and La$_4$Ni$_3$O$_{10}$ under pressure appears reasonable because the oxygen anions are likely to shrink more under pressure than the cations. Assuming the pressure dependence of $r_{\mathrm{Ni}}$ and $r_{\mathrm{La}}$ is negligible, the derivative of the tolerance factor with respect to the oxygen ionic radius, 
$\partial t/\partial r_{\mathrm{O}} = (r_{\mathrm{Ni}} - r_{\mathrm{La}})/\sqrt{2}(r_{\mathrm{Ni}} + r_{\mathrm{O}})^2$, 
is negative because $r_{\mathrm{Ni}} < r_{\mathrm{La}}$.

\subsection{Isovalent Substitution of La Ions -- Chemical Pressure --}

As mentioned above, superconductivity in La$_3$Ni$_2$O$_7$ and La$_4$Ni$_3$O$_{10}$ occurs when the compounds adopt the tetragonal $I4/mmm$ symmetry. 
This is consistent with theoretical expectations that the hopping integral between two Ni ions along the $c$-axis, $t_{\perp}$, is crucial for the emergence of superconductivity \cite{NakataPRB2017}.  
In this context, one might consider that substitution of La ions with smaller tetravalent ions could induce chemical pressure, removing the orthorhombic or monoclinic distortion and enabling superconductivity at ambient pressure.  
However, this simple idea cannot be applied. 
The orthorhombicity, $b/a$, increases when smaller ions such as Pr or Nd substitute La, as shown in Table \ref{LP}, reflecting the reduction of the tolerance factor for smaller ions. 
This trend has been confirmed in systematic experiments \cite{WangnpjQM2025,FrugierJMCA2025}. 
Consequently, for smaller $R$ ions, higher pressures are required to eliminate the distortion, as suggested by theoretical calculations for La$_3$Ni$_2$O$_7$ \cite{Geislernpj2024} and supported by experimental observations \cite{WangnpjQM2025}. 
This illustrates that internal chemical pressure and external applied pressure can have different effects on the structural transition.

There are two distinct La sites in La$_3$Ni$_2$O$_7$ and La$_4$Ni$_3$O$_{10}$: La(1) in the perovskite slabs, and La(2) in the rock-salt slabs. 
Substitution of La ions with smaller ions may affect the transition pressure differently depending on which site is predominantly substituted. 
For example, if the La(2) sites were fully substituted while the La(1) sites remained unchanged, the perovskite slabs would be compressed in the $ab$ plane relative to the pure La compounds, which could reduce the transition pressure. 
From this perspective, it is noteworthy that a sample of La$_2$PrNi$_2$O$_7$ showed a lower transition pressure of 11 GPa compared with 14 GPa for La$_3$Ni$_2$O$_7$ \cite{WangNature2024}, despite the smaller average ionic radius of the La/Pr sites relative to pure La, although the site occupancy of Pr ions is not reported. 
Since the site occupancy of La and substituted ions depends on synthesis conditions, such as cooling rate and oxygen partial pressure, the transition pressure may be tunable by adjusting these conditions.

\begin{table}
\caption{\label{LP} Lattice parameters of La$_3$Ni$_2$O$_7$ with orthorhombic $Amam$ structure, La$_4$Ni$_3$O$_{10}$ with monoclinic $P2_1/a$ structure, and their La-substituted materials at room temperature and ambient pressure. The oxygen contents of some materials are nominal. $*$: $Fmmm$ was tentatively used to estimate the lattice parameters.}
\resizebox{\textwidth}{!}{
\begin{tabular}{l|c|c|c|c|c|c|c}
Composition & Space Group & $a$ (\AA) & $b$ (\AA)& $c$ (\AA) & $\beta$ ($^{\circ}$) & $b/a$ & Ref. \\ \hline
La$_3$Ni$_2$O$_{7.02}$ & $Amam$ & 5.39283 & 5.44856 & 20.5185 & -- & 1.01033 & \cite{LingJSSC1999} \\
La$_3$Ni$_2$O$_{7.05}$ & $Amam$ & 5.39710 & 5.45011 & 20.5074 & -- & 1.00982 & \cite{LingJSSC1999} \\
La$_3$Ni$_2$O$_7$ & $Amam$ & 5.3981 & 5.4494 & 20.502 & -- & 1.0095 & \cite{VoroninNIMPRA2001} \\
La$_3$Ni$_2$O$_7$ & $Amam$ & 5.392 & 5.447 & 20.517 & -- & 1.010 & \cite{DeminaInorgMater2005} \\
La$_3$Ni$_2$O$_7$ & $Amam$ & 5.3959 & 5.4495 & 20.5350 & -- & 1.0099 & \cite{KiselevInorgMater2007} \\
La$_3$Ni$_2$O$_{7.01}$ & $Amam$ & 5.4000 & 5.4384 & 20.455 & -- & 1.0071 & \cite{LiuSciChinaPhysMechAstro2022} \\
La$_3$Ni$_2$O$_{6.93}$ & $Amam$ & 5.3920 & 5.4480 & 20.5311 & -- & 1.0104 & \cite{WangPRX2024} \\
La$_3$Ni$_2$O$_{6.93}$ & $Amam$ & 5.38996 & 5.44719 & 20.5305 & -- & 1.01062 & \cite{WangInorgChem2025} \\
La$_3$Ni$_2$O$_{7.01}$ & $Amam$ & 5.3903 & 5.4464 & 20.507 & -- & 1.0104 & \cite{UekiJPSJ2025} \\
La$_3$Ni$_2$O$_7$ & $Amam$ & 5.392 & 5.447 & 20.517 & -- & 1.010 & \cite{DeminaInorgMater2005} \\
La$_3$Ni$_2$O$_{6.92}$ & $Amam$ & 5.4018 & 5.4557 & 20.537 & -- & 1.0100 & \cite{ChenPRL2024} \\
La$_3$Ni$_2$O$_7$ & $Amam$ & 5.39885 & 5.45130 & 20.5163 & -- & 1.00972 & \cite{Zhang.arXiv.2025.01501} \\
La$_3$Ni$_2$O$_7$ & $Amam$ & 5.407 & 5.4176 & 20.490 & -- & 1.002 & \cite{Huo.arXiv.2501.15929} \\ \hline

La$_2$PrNi$_2$O$_{7.01}$ & $Amam$ & 5.37318 & 5.45274 & 20.41213 & -- & 1.01481 & \cite{WangNature2024} \\ \hline

La$_4$Ni$_3$O$_{10}$ & $P2_1/a$ & 5.4151 & 5.4714 & 14.2277 & 100.818 & 1.0104 & \cite{ZhangPRM2020} \\
La$_4$Ni$_3$O$_{10}$ & $P2_1/a$ & 5.4160 & 5.4656 & 27.9750 & 90.179 & 1.0092 & \cite{SongJMaterChemA2020} \\
La$_4$Ni$_3$O$_{10}$ & $P2_1/a$ & 5.4243 & 5.4748 & 28.0053 & 90.192 & 1.0093 & \cite{RoutPRB2020} \\
La$_4$Ni$_3$O$_{10}$ & $P2_1/a$ & 5.42335 & 5.47315 & 28.00414 & 90.1475 & 1.00918 & \cite{KumarJMMM2020} \\
La$_4$Ni$_3$O$_{10.03}$ & $P2_1/a$ & 5.403 & 5.454 & 14.24 & 100.70 & 1.009 & \cite{HuangfuPRR2020} \\
La$_4$Ni$_3$O$_{10}$ & $P2_1/a$ & 5.4142 & 5.4647 & 14.2208 & 100.66 & 1.0093 & \cite{ZhuNature2024} \\
La$_4$Ni$_3$O$_{10}$ & $P2_1/a$ & 5.4164 & 5.4675 & 14.2279 & 100.752 & 1.0094 & \cite{LiSCPMA2024} \\
La$_4$Ni$_3$O$_{10}$ & $P2_1/a$ & 5.4247 & 5.4596 & 14.2422 & 100.9663 & 1.0064 & \cite{LiCrystGrowthDesign2024} \\
La$_4$Ni$_3$O$_{10}$ & $P2_1/a$ & 5.4132 & 5.4627 & 14.2337 & 100.71 & 1.0091 & \cite{LiChiPhysLett2024} \\
La$_4$Ni$_3$O$_{10}$ & $P2_1/a$ & 5.4131 & 5.4638 & 14.2461 & 101.26 & 1.0094 & \cite{LiChiPhysLett2024} \\
La$_4$Ni$_3$O$_{10}$ & $P2_1/a$ & 5.4082 & 5.4533 & 14.2548 & 100.94 & 1.0083 & \cite{LiChiPhysLett2024} \\
La$_4$Ni$_3$O$_{10}$ & $P2_1/a$ & 5.4184 & 5.4655 & 14.2278 & 100.841 & 1.0087 & \cite{Shi.arXiv2501.12647v1} \\
La$_4$Ni$_3$O$_{10}$ & $P2_1/a$ & 5.4185 & 5.4678 & 14.2296 & 100.7245 & 1.0091 & \cite{Li.arXiv.2501.13511v1} \\
La$_4$Ni$_3$O$_{9.99}$ & $P2_1/a$ & 5.4162 & 5.4642 & 27.984 & 90.256 & 1.0089 & \cite{Khasanov.arXiv.2503.04400v1} \\ \hline

Pr$_4$Ni$_3$O$_{10}$ & $Fmmm^{*}$ & 5.370 & 5.462 & 27.528 & -- & 1.017 & \cite{ZhangJSSC1995} \\
Pr$_4$Ni$_3$O$_{10}$ & $Fmmm^{*}$ & 5.372 & 5.462 & 27.532 & -- & 1.017 & \cite{BassatEurJSSIC1998} \\
Pr$_4$Ni$_3$O$_{10.06}$ & $Fmmm^{*}$ & 5.369 & 5.464 & 27.523 & -- & 1.018 & \cite{BassatEurJSSIC1998} \\
Pr$_4$Ni$_3$O$_{10.1}$ & $Fmmm^{*}$ & 5.356 & 5.463 & 27.548 & -- & 1.020 & \cite{BassatEurJSSIC1998} \\
Pr$_4$Ni$_3$O$_{10}$ & $Fmmm$ & 5.3714 & 5.4611 & 27.5271 & -- & 1.0167 & \cite{VibhuJPS2016} \\
Pr$_4$Ni$_3$O$_{10}$ & $P2_1/a$ & 5.3816 & 5.4711 & 14.0284 & 100.646 & 1.0166 & \cite{ZhangPRM2020} \\
Pr$_4$Ni$_3$O$_{10.1}$ & $P2_1/a$ & 5.3705 & 5.4637 & 27.5728 & 90.303 & 1.0174 & \cite{TsaiJSSC2020} \\
Pr$_4$Ni$_3$O$_{10.1}$ & $P2_1/a$ & 5.37556 & 5.46462 & 27.5463 & 90.283 & 1.01657 & \cite{SongJMaterChemA2020} \\
Pr$_4$Ni$_3$O$_{10}$ & $P2_1/a$ & 5.3826 & 5.4717 & 27.583 & 90.284 & 1.0166 & \cite{RoutPRB2020} \\
Pr$_4$Ni$_3$O$_{9.97}$ & $P2_1/a$ & 5.372 & 5.458 & 14.02 & 100.81 & 1.016 & \cite{HuangfuPRR2020} \\ \hline

Nd$_4$Ni$_3$O$_{10}$ & $Fmmm^{*}$ & 5.362 & 5.454 & 27.410 & -- & 1.017 & \cite{ZhangJSSC1995} \\
Nd$_4$Ni$_3$O$_{10.1}$ & $P2_1/a$ & 5.36373 & 5.45221 & 27.4100 & 90.292 & 1.01650 & \cite{SongJMaterChemA2020} \\
Nd$_4$Ni$_3$O$_{10}$ & $P2_1/a$ & 5.3719 & 5.46 & 27.4506 & 90.299 & 1.0164 & \cite{RoutPRB2020} \\
Nd$_4$Ni$_3$O$_{9.93}$ & $P2_1/a$ & 5.351 & 5.441 & 13.93 & 100.79 & 1.017 & \cite{HuangfuPRR2020} \\ \hline

La$_3$PrNi$_3$O$_{10.1}$ & $P2_1/a$ & 5.396 & 5.458 & 14.17 & 100.71 & 1.011 & \cite{HuangfuPRR2020} \\
La$_2$Pr$_2$Ni$_3$O$_{10.01}$ & $P2_1/a$ & 5.382 & 5.457 & 14.12 & 100.71 & 1.014 & \cite{HuangfuPRR2020} \\
LaPr$_3$Ni$_3$O$_{10}$ & $P2_1/a$ & 5.382 & 5.466 & 14.07 & 100.78 & 1.016 & \cite{HuangfuPRR2020} \\

La$_3$NdNi$_3$O$_{9.95}$ & $P2_1/a$ & 5.401 & 5.466 & 14.19 & 100.68 & 1.012 & \cite{HuangfuPRR2020} \\
La$_2$Nd$_2$Ni$_3$O$_{9.99}$ & $P2_1/a$ & 5.375 & 5.463 & 14.14 & 100.67 & 1.016 & \cite{HuangfuPRR2020} \\
LaNd$_3$Ni$_3$O$_{9.98}$ & $P2_1/a$ & 5.363 & 5.459 & 14.08 & 100.73 & 1.018 & \cite{HuangfuPRR2020} \\

Pr$_3$NdNi$_3$O$_{9.95}$ & $P2_1/a$ & 5.361 & 5.451 & 13.98 & 100.79 & 1.017 & \cite{HuangfuPRR2020} \\
Pr$_2$Nd$_2$Ni$_3$O$_{9.93}$ & $P2_1/a$ & 5.358 & 5.449 & 13.96 & 100.78 & 1.017 & \cite{HuangfuPRR2020} \\
PrNd$_3$Ni$_3$O$_{10.07}$ & $P2_1/a$ & 5.356 & 5.447 & 13.95 & 100.80 & 1.017 & \cite{HuangfuPRR2020} \\

\end{tabular}
}
\end{table}

\subsection{Oxygen Nonstoichiometry}

The crystal structures of La$_3$Ni$_2$O$_7$ and La$_4$Ni$_3$O$_{10}$ are not well defined as mentioned above, and their physical properties are sample-dependent, as shown later. 
In such cases, oxygen content is often the primary cause of the sample dependence. 
Indeed, the oxygen content of Ni oxides can vary depending on synthesis conditions. Here, oxygen defects are discussed from the viewpoint of crystal structures, while their precise control will be addressed in a later section. 
A deficiency of La and/or Ni ions may also be possible, considering perovskite-related materials such as ReO$_3$, La$_{2/3}$TiO$_3$ \cite{AbeMRB1974}, and La$_{1-\epsilon}$Mn$_{1-\epsilon}$O$_3$ \cite{TopferJSSC1997}. 
However, no reports on cation deficiency in La$_3$Ni$_2$O$_7$ or La$_4$Ni$_3$O$_{10}$ are available.

Oxygen deficiency readily occurs in the perovskite slabs. 
It is well known that perovskite oxides, $AM$O$_3$, can accommodate oxygen loss. 
When 0.5 oxygen atoms per formula unit are systematically removed, brownmillerite-type structures appear, as in Sr$_2$Fe$_2$O$_5$ \cite{GallagherJCP1964}. 
The crystal structure of the infinite-layer nickelate superconductor is also derived from perovskite Ni oxides \cite{LiNature2019}. 
La$_3$Ni$_2$O$_6$ and La$_4$Ni$_3$O$_8$ can be obtained by nearly topotactic removal of oxygen atoms located between two Ni atoms along the $c$-axis (inner apical oxygen), although the rock-salt slabs between perovskite layers transform into fluorite-type structures in both compounds \cite{LacorreJSSC1992,PoltavetsJACS2006,PoltavetsInorgChem2007}. 
The critical oxygen content required to preserve the rock-salt structure remains unclear. 
However, since the fluorite structure in La$_4$Ni$_3$O$_8$ reverts to the rock-salt structure above 21 GPa \cite{ChengPRL2012}, the enthalpy gain associated with the rock-salt to fluorite transformation is likely small. 
These facts suggest that oxygen deficiency predominantly occurs in the perovskite slabs, specifically at the inner apical oxygen sites. 
Indeed, oxygen loss at these sites has been directly observed in La$_3$Ni$_2$O$_7$ by neutron diffraction \cite{PoltavetsMRB2006} and multislice electron ptychography \cite{DongNature2024}. 
Upon oxygen loss, adjacent NiO$_5$ pyramids face each other base-to-base. 
Since this structural motif is also found in other compounds such as YBa$_2$Cu$_3$O$_{7-\delta}$, this type of oxygen deficiency appears to be a common feature of RP phases with $n \neq 1$.

\begin{figure}
\includegraphics[width=8cm]{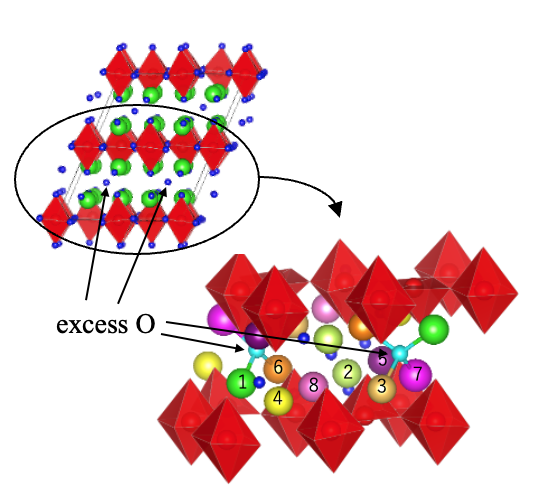}
\caption{\label{ExcessO} 
Local structure of La$_2$NiO$_{4.25}$ \cite{DemourguesJSSC1993}. 
Numbered circles represent La atoms at their respective sites, while blue and pink atoms denote excess and original oxygen atoms, respectively.
}
\end{figure}

No detailed information has been reported regarding the incorporation of excess oxygen in La$_3$Ni$_2$O$_7$ and La$_4$Ni$_3$O$_{10}$. 
However, detailed studies of La$_2$NiO$_4$ \cite{JorgensenPRB1989,HiroiPRB1990} strongly suggest that excess oxygen is located within the rock-salt slabs. 
The rock-salt slabs of La$_2$CuO$_4$ also accommodate excess oxygen \cite{JorgensenPRB1988}, supporting this assignment. 
Thus, if excess oxygen exists in La$_3$Ni$_2$O$_7$ and La$_4$Ni$_3$O$_{10}$, it is also most likely located in the rock-salt slabs. 
For La$_2$NiO$_{4+\delta}$, excess oxygen atoms form commensurate ordered structures at $\delta = 1/4$ (0.25), $1/6$ (0.17), $1/8$ (0.13), and so on \cite{HiroiPRB1990}. 
In La$_2$NiO$_{4.25}$, the introduction of excess oxygen induces a strong distortion of the crystal structure from orthorhombic $Bmab$ to monoclinic $C2$ \cite{DemourguesJSSC1993}. 
In this phase, four of the eight La sites with Wyckoff position $4c$ in the rock-salt slabs are located adjacent to an excess oxygen atom, as illustrated in Fig.~\ref{ExcessO}. 
Because excess oxygen in the rock-salt slabs modifies the tilting pattern of NiO$_6$ octahedra \cite{TranquadaPRB1994}, it significantly alters the crystal fields at the Ni sites. 
Thus, excess oxygen in La$_3$Ni$_2$O$_7$ and La$_4$Ni$_3$O$_{10}$ can strongly influence physical properties beyond the rigid-band picture.

The kinetics of excess oxygen atoms is also important. 
RP phases of nickel oxides exhibit such high oxygen-ion conductivity that they are considered as potential cathode materials for solid oxide fuel cells (SOFCs) \cite{YatooMaterialsToday2022}. 
Thus, even if homogeneous samples are obtained at high temperatures, they may undergo order–disorder transitions of excess oxygen atoms upon cooling, leading to phase separation. 
Indeed, La$_2$NiO$_{4+\delta}$ exhibits multiple phases below room temperature \cite{JorgensenPRB1989,TranquadaNature1995,KyomenPRB1995,KyomenPRB1999}, indicating that oxygen content—and hence band filling—cannot be controlled continuously and homogeneously at ambient temperatures. 
Furthermore, the order–disorder transitions are slow \cite{KyomenPRB1995,KyomenPRB1999}, potentially reducing reproducibility of experimental results unless special care is taken when excess oxygen is present in the rock-salt slabs. 
Notably, phase separation also occurs in single crystals.

As mentioned above, La$_3$Ni$_2$O$_7$ and La$_4$Ni$_3$O$_{10}$ can accommodate both oxygen deficiency and excess oxygen. 
Thus, even nominally stoichiometric or oxygen-deficient samples may contain excess oxygen in the rock-salt slabs, because oxygen atoms located between Ni ions along the $c$-axis can easily migrate to the rock-salt layers. 
This type of defect is widely recognized in solids and is known as a Frenkel defect. 
In short, La$_3$Ni$_2$O$_7$ and La$_4$Ni$_3$O$_{10}$ have a strong tendency to form oxygen Frenkel defects. 
Interestingly, nearly stoichiometric La$_3$Ni$_2$O$_7$ exhibits two NQR (nuclear quadrupole resonance) peaks for La(2) sites, indicating two distinct local environments \cite{YashimaJPSJ2025}. 
The smaller peak has been attributed to La(2) sites adjacent to oxygen vacancies between Ni ions along the $c$-axis. 
From the relative intensities of the two peaks, the oxygen deficiency was estimated to be approximately 1/8, which coincides with the $\delta$ value that induces a commensurate superstructure in La$_2$NiO$_{4.17}$, despite the nearly stoichiometric oxygen content of the sample (7.01) \cite{UekiJPSJ2025}. 
This suggests that approximately one-eighth of the oxygen atoms migrate to the rock-salt slabs as Frenkel defects. 
Notably, even single crystals contain Frenkel defects—likely in greater abundance than powders—since they are typically synthesized at much higher temperatures. 
Because excess oxygen in the rock-salt slabs can induce phase separation, the small superconducting volume fractions discussed below may originate from phase separation caused by such Frenkel defects.

\subsection{Stacking Faults and Related Materials}

Stacking faults have been studied more extensively and are often considered more significant than Frenkel defects. 
RP-phase compounds usually contain stacking faults because of the structural similarity among phases with different $n$, which likely possess comparable formation enthalpies. 
High-resolution transmission electron microscopy clearly reveals stacking faults in La$_3$Ni$_2$O$_7$ and La$_4$Ni$_3$O$_{10}$ \cite{CarvalhoJMC1997,DrennanMRB1982,RamJSSC1986,SreedharJSSC1994}, although their quantitative evaluation remains difficult. 
These common defects were once regarded as highly relevant to the superconductivity of La$_3$Ni$_2$O$_7$, leading to the proposal that superconductivity might be filamentary, arising at the interfaces with La$_4$Ni$_3$O$_{10}$ inclusions, present as stacking faults, rather than being a bulk property \cite{Zhou.arXiv.2311.12361v1}. 
This view was supported by the small superconducting volume fraction estimated from AC susceptibility \cite{ZhouMRE2025}. 
However, this scenario is now less convincing, as large superconducting volume fractions of $\sim41$\% at 22.0 GPa were reported for La$_3$Ni$_2$O$_7$ \cite{LiNatSciRev2025}, and La$_2$PrNi$_2$O$_7$ also exhibits large values of approximately 57\% at 20 GPa and 97\% at 19 GPa \cite{WangNature2024}. 

\begin{figure}
\includegraphics[width=5cm]{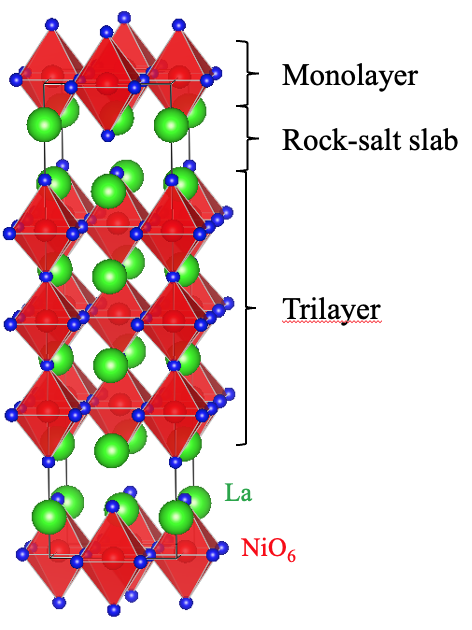}
\caption{\label{1313phase} 
Crystal structure of the newly discovered 1313 phase of lanthanum nickelate.
}
\end{figure}

Interestingly, a new polymorph of La$_3$Ni$_2$O$_7$ has been discovered \cite{PuphalPRL2024,ChenJACS2024,WangInorgChem2025}, which is related to stacking-fault–like structural motifs. 
The conventional La$_3$Ni$_2$O$_7$ is composed of alternating double perovskite slabs and single rock-salt slabs, as described earlier. 
By contrast, the new La$_3$Ni$_2$O$_7$ consists of alternating single and triple perovskite slabs, replacing the double perovskite slabs of the conventional structure, as illustrated in Fig.~\ref{1313phase}. 
In short, it is a chimera of the conventional $n=1$ and 3 phases, and it has been named the 1313 phase to distinguish it from the conventional La$_3$Ni$_2$O$_7$, now referred to as the 2222 phase. 
The space group of the 1313 phase is reported as orthorhombic $Fmmm$ (\#69) \cite{PuphalPRL2024} or $Cmmm$ (\#65) \cite{ChenJACS2024,WangInorgChem2025} at ambient pressure. 
In either case, the Ni--O--Ni bridging angles along the $c$-axis in the $n=3$ perovskite slabs deviate from 180$^{\circ}$. 
For the $Fmmm$ structure, this angle increases gradually under pressure above $\sim$6 GPa and reaches 180$^{\circ}$ via a structural transition at 12.3 GPa \cite{PuphalPRL2024}. 
Superconductivity then emerges above $\sim$8 GPa, with transition temperatures comparable to those of the 2222 phase. 
For the $Cmmm$ structure, no high-pressure measurements have yet been reported, and thus it remains uncertain whether superconductivity occurs. 
At ambient pressure, one sample shows semiconducting behavior \cite{WangInorgChem2025}, while another displays metallic behavior with an anomaly in resistivity near 134 K, likely due to density-wave formation, as observed in the 2222 phase \cite{ChengPRL2012}. 
Oxygen nonstoichiometry, Frenkel defects, stacking faults, and other sample-dependent factors are, of course, also expected to influence the physical properties of the 1313 phase.

The discovery of polymorphism in La$_3$Ni$_2$O$_7$ broadens the landscape of RP-related materials. 
For example, shortly thereafter, La$_2$NiO$_4\cdot$La$_3$Ni$_2$O$_7$ (= La$_5$Ni$_3$O$_{11}$) was reported to adopt a chimera structure combining the $n=1$ and $n=2$ phases \cite{LiPRM2024}. 
This compound no longer follows the general RP chemical formula A$_{n+1}$M$_{n}$X$_{3n+1}$. 
The 1212 phase crystallizes in the orthorhombic $Immm$ (\#71) structure at ambient pressure, where the Ni--O--Ni bridging angle along the $c$-axis in the $n=2$ perovskite slabs is 180$^{\circ}$. 
Its resistivity shows a metal–semiconductor crossover with temperature, but no superconductivity has been observed in either resistivity or magnetic susceptibility measurements on as-grown single crystals \cite{LiPRM2024}. 
In this review, La$_3$Ni$_2$O$_7$ refers to the 2222 phase unless otherwise specified.

\section{Synthesis \& Characterization}

\subsection{Powder Synthesis}
La$_3$Ni$_2$O$_7$ and La$_4$Ni$_3$O$_{10}$ were synthesized intentionally for the first time in 1981 \cite{BrisiJLCM1981} and 1979 \cite{SeppanenSJM1979}, respectively, although these phases had previously been recognized as impurities in synthetic LaNiO$_3$ or La$_2$NiO$_4$. 
Since then, many attempts have been made to synthesize high-quality samples, especially after they were identified as promising SOFC cathode materials in 2006 \cite{AmowSSI2006RFR14,YatooMaterialsToday2022}. 
For most syntheses, the Pechini method (a sol--gel process) has been employed, although the number of reports is too large to be fully summarized here. 
In the solution-based process, atoms are homogeneously dispersed in a gel when gelation is well controlled, drastically facilitating the approach to equilibrium.
Although the details of sol--gel processes vary among reports, the representative Pechini method \cite{AmowSSI2006RFR14} is briefly introduced here. 
A stoichiometric mixture of La(NO$_3$)$_3$ and Ni(NO$_3$)$_2$ is dissolved in water, to which excess citric acid and ethylene glycol are added. 
The solution is then dried to form a gel, which is subsequently heated at 750$^{\circ}$C in air to remove organic components. 
The resulting powder is pressed into pellets and fired at 1100$^{\circ}$C for La$_3$Ni$_2$O$_7$ or 1050$^{\circ}$C for La$_4$Ni$_3$O$_{10}$ for several days. 
Another frequently used method is the solid-state reaction of nitrates. 
Compared with simple oxides, nitrates are unstable and their powders are usually fine, which enhances the reaction rate. 
However, even with nitrates, obtaining homogeneous samples is difficult \cite{BrisiJLCM1981,RamJSSC1986}. 
These findings suggest that direct solid-state reactions from La$_2$O$_3$ and NiO powders do not proceed smoothly to equilibrium. 
In fact, several reports explicitly state that single-phase samples could not be obtained from such reactions \cite{ZhangJSSC1994n1,LingJSSC1999,VoroninNIMPRA2001}. 

Samples prepared from stoichiometric mixtures of La$_2$O$_3$ and NiO generally contain significant amounts of other RP phases \cite{SakuraiSSP2025}. 
The large intensity of secondary-phase peaks in the sample fired at 1200$^{\circ}$C suggests that high-quality samples cannot be obtained, even after repeated grinding and firing intended to produce ``single-phase'' products. 
Nevertheless, the synthesis temperature cannot be raised significantly, as only a narrow temperature window is available for each compound.
In the samples fired at 1000$^{\circ}$C, the raw materials of La$_2$O$_3$ and NiO remained, suggesting that the reaction temperature is too low. 
By contrast, La$_3$Ni$_2$O$_7$ decomposes into La$_2$NiO$_4$ and NiO above 1366$^{\circ}$C, and La$_4$Ni$_3$O$_{10}$ decomposes into La$_3$Ni$_2$O$_7$ and NiO above 1277$^{\circ}$C at an oxygen partial pressure of $p_{\mathrm{O_2}}=1$ bar \cite{AdachiJAmCerSoc2019}.

The combination of ball milling and CIP (cold isostatic pressing) has been reported to be effective for synthesizing La$_3$Ni$_2$O$_7$ and La$_4$Ni$_3$O$_{10}$ samples suitable for SOFC applications from La$_2$O$_3$ and NiO powders \cite{TakahashiJAmCeramSoc2010}, although it remains uncertain whether the quality is sufficient for microscopic measurements of their electronic properties. 
A La$_3$Ni$_2$O$_{6.97}$ sample for SOFC studies, prepared either from ball-milled mixtures of La$_2$O$_3$ and NiO or from nitrates \cite{AdachiJAmCerSoc2019}, exhibited significant NQR (nuclear quadrupole resonance) signals arising from La$_4$Ni$_3$O$_{10}$ intergrowth \cite{YashimaJPSJ2025}. 

\begin{figure}
\includegraphics[width=8cm]{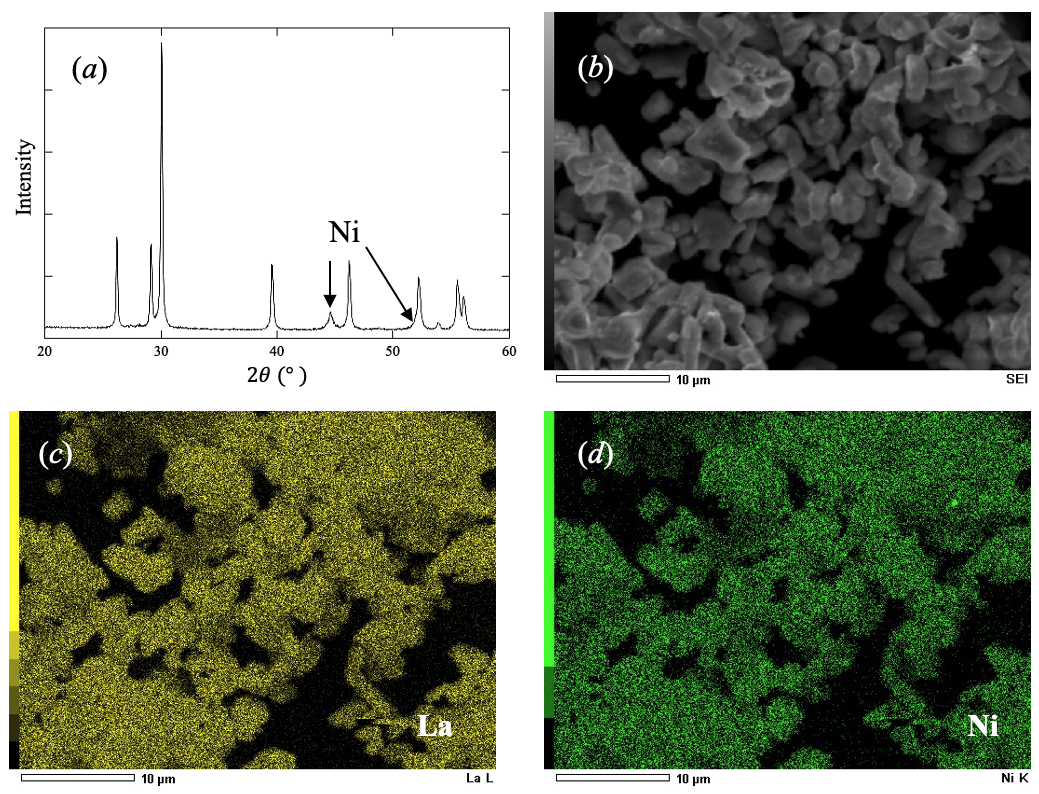}
\caption{\label{La2O3andNi} 
XRD pattern ($a$), SEM (scanning electron microscopy) image ($b$), and EDX (energy-dispersive X-ray spectroscopy) mapping of La ($c$) and Ni ($d$) in the sample reduced by hydrogen gas after preliminary synthesis.
}
\end{figure}

Recently, a new method was developed to synthesize homogeneous La$_3$Ni$_2$O$_7$ and La$_4$Ni$_3$O$_{10}$ directly from La$_2$O$_3$ and NiO \cite{SakakibaraPRB2024,SakuraiSSP2025}. 
In this approach, the samples are preliminarily heated at 1300$^{\circ}$C (for La$_3$Ni$_2$O$_7$) or 1200$^{\circ}$C (for La$_4$Ni$_3$O$_{10}$), and then reduced under flowing 10\% H$_2$/Ar gas to yield a mixture of La$_2$O$_3$ and Ni metal. 
This method was originally developed for the synthesis of Y$_2$Ba$_4$Cu$_7$O$_{15-x}$ to obtain more homogeneous samples than those produced by the polymerized-complex method (a type of solution process) \cite{KatoJSSC1998}. 
As shown in Fig.~\ref{La2O3andNi}$a$, the XRD peaks of Ni metal in a reduced sample are broad, indicating that the Ni particles are very fine. 
Indeed, Ni atoms are dispersed within individual micron-sized grains, as observed in Fig.~\ref{La2O3andNi}$b$--$d$. 
The reduced samples are subsequently fired twice at 1300$^{\circ}$C for La$_3$Ni$_2$O$_7$ or 1200$^{\circ}$C for La$_4$Ni$_3$O$_{10}$, yielding single-phase products \cite{SakakibaraPRB2024,UekiJPSJ2025,YashimaJPSJ2025}. 
La$_3$Ni$_2$O$_7$ synthesized by this method exhibits no NQR signals from La$_4$Ni$_3$O$_{10}$ \cite{YashimaJPSJ2025}.
By contrast, La$_4$Ni$_3$O$_{10}$ samples still contain intergrowth of La$_3$Ni$_2$O$_7$, as evidenced by NQR signals from the latter \cite{YashimaJPSJ2025}. 
Indeed, the resistivity of La$_4$Ni$_3$O$_{10}$ under high pressure shows a drop below $\sim$80~K, probably due to the La$_3$Ni$_2$O$_7$ intergrowth. 
To suppress this, HIP (hot isostatic pressing) annealing at 1200$^{\circ}$C under high oxygen pressure ($p_{\mathrm{O_2}}=400$~atm) is effective, because La$_3$Ni$_2$O$_7$ decomposes into La$_4$Ni$_3$O$_{10}$ and La$_2$O$_3$ under such conditions \cite{AdachiJAmCerSoc2019}.

\subsection{Single Crystal}
Single crystals of La$_3$Ni$_2$O$_7$ and La$_4$Ni$_3$O$_{10}$ are grown by the FZ (floating zone) and flux methods. 
These compounds exhibit incongruent melting at $p_{\mathrm{O_2}}=1$ bar \cite{AdachiJAmCerSoc2019} as mentioned above; La$_{n+1}$Ni$_n$O$_{3n+1}$ with larger $n$ decomposes stepwise into phases with smaller $n$ as temperature increases. 
The Ni valence for RP phases with larger $n$ is higher, so at higher temperatures oxygen gas is released from the oxides due to the larger entropy of the gas phase. 
Fortunately, there exist oxygen partial pressures that allow congruent melting, although their exact melting points have not been reported. 
These pressures are approximately 14 bar for La$_3$Ni$_2$O$_7$, 16–30 bar for La$_4$Ni$_3$O$_{10}$, and above 50 bar for LaNiO$_3$ \cite{ZhangPRM2020}. 
Thus, the FZ growth of single crystals is performed at 15 bar for La$_3$Ni$_2$O$_7$ \cite{ZhangNatPhys2024,WangInorgChem2025,SunNature2023,PuphalPRL2024,LiuSciChinaPhysMechAstro2022,Huo.arXiv.2501.15929,DongNature2024,ChenPRB2025} and 20 bar for La$_4$Ni$_3$O$_{10}$ \cite{YuanJCG2024,ZhuNature2024,ZhangPRM2020,ZhangPRX2025,LiSCPMA2024}. 
No information on the melt temperature, except 1650$^{\circ}$C for the growth of La$_4$Ni$_3$O$_{10}$ \cite{YuanJCG2024}, has been reported, but high-power Xenon arc lamps were used. 
Since the crystal growth of Pr$_4$Ni$_3$O$_{10}$ requires $p_{\mathrm{O_2}}>100$ bar \cite{ZhangPRM2020}, the synthesis of La-substituted crystals likely requires larger $p_{\mathrm{O_2}}$ than that of the non-doped crystals.

For crystal growth by the FZ method, the occurrence of stacking faults is almost inevitable, especially for La$_3$Ni$_2$O$_7$, because the samples pass through the temperature window in which La$_4$Ni$_3$O$_{10}$ is stable at high $p_{\mathrm{O_2}}$ while being cooled from the melts. 
Indeed, stacking faults were clearly observed by TEM \cite{ZhouMRE2025}, and some unidentified peaks were observed in $^{139}$La NMR signals \cite{ZhaoSciBull2025}.
This seems to be a major reason why the 1313 phase was obtained by the FZ method \cite{ChenJACS2024,PuphalPRL2024,WangInorgChem2025}, as supported by annealing powder samples at various $p_{\mathrm{O_2}}$ \cite{Zhang.arXiv.2025.01501}, although this is unlikely the only reason, considering that the 1212 phase was synthesized by the flux method \cite{LiPRM2024}. 
It is unclear whether the stacking faults can be eliminated by annealing the crystals at low oxygen pressures, such as $p_{\mathrm{O_2}}=1$ atm at 1300$^{\circ}$C (where La$_4$Ni$_3$O$_{10}$ decomposes into La$_3$Ni$_2$O$_7$ and NiO, as mentioned above). 
By contrast, La$_4$Ni$_3$O$_{10}$ crystals have stacking faults of double perovskite slabs in some cases \cite{YuanJCG2024}, although they may disappear by high $p_{\mathrm{O_2}}$ and temperature annealing, such as HIP treatment. 
As-cast La$_4$Ni$_3$O$_{10}$ crystals can have the orthorhombic $Bmeb$ structure, which changes into the monoclinic $P2_1/a$ structure upon annealing, supporting that the $Bmeb$ structure is metastable \cite{ZhangPRM2020,YuanJCG2024}.

Unlike the FZ method, the flux method does not require very high temperatures or high $p_{\mathrm{O_2}}$, although it carries the risk of inclusions in the grown crystals and side reactions during flux removal. 
For successful synthesis, it is important to choose proper flux materials, although no systematic principle exists. 
Several groups have succeeded in this regard, and according to their reports, K$_2$CO$_3$ or a NaCl/KCl mixture works well as a flux at 1050$^{\circ}$C or below for La$_3$Ni$_2$O$_7$ \cite{Li.arXiv.2501.14584,ShiarXiv.2501.14202}, whereas only K$_2$CO$_3$ flux has been reported for La$_4$Ni$_3$O$_{10}$ to date \cite{Li.arXiv.2501.13511v1,LiCrystGrowthDesign2024,Shi.arXiv2501.12647v1}. 
The crystals are almost square or rectangular plates with dimensions of $\sim100^2$ $\mu$m$^2$ for both compounds \cite{Li.arXiv.2501.14584,Li.arXiv.2501.13511v1}.

Some single crystals grown by the flux methods sometimes have different crystallographic symmetries from orthorhombic $Amam$, $Fmmm$, $Bmeb$, and monoclinic $P2_1/a$. 
La$_3$Ni$_2$O$_7$ crystals made by the K$_2$CO$_3$ flux method adopt the monoclinic $P2_1/m$ structure \cite{Li.arXiv.2501.14584}, whereas those made by the NaCl/KCl flux method adopt the orthorhombic $Amam$ structure \cite{ShiarXiv.2501.14202}. 
The as-cast crystals by the latter method were electrically insulating, and after slow cooling from 500$^{\circ}$C to 50$^{\circ}$C at 10–15 bar or 150 bar oxygen gas, orthorhombic $Amam$ and tetragonal $I4/mmm$ crystals showing metallic behavior were obtained. 
Although the oxygen content of the tetragonal crystals was estimated to be 6.96 by X-ray structural analysis, it most likely exceeds 7. 
La$_3$Ni$_2$O$_7$ transforms from the orthorhombic to the tetragonal structure upon annealing at extremely high $p_{\mathrm{O_2}}$, as mentioned later. 
The underestimation is probably due to the assumption of no excess oxygen atoms in the rock-salt slabs. 
On the other hand, tetragonal La$_4$Ni$_3$O$_{10}$ crystals with $I4/mmm$ symmetry were obtained directly by the flux method in flowing oxygen gas. 
Since monoclinic $P2_1/a$ crystals were obtained from flux growth in air, the tetragonal structure of La$_4$Ni$_3$O$_{10}$ is likely caused by excess oxygen atoms in the rock-salt slabs, suggesting that the flux method is effective for introducing excess oxygen atoms.

The flux methods are also applicable to La$_{3-x}R_x$Ni$_2$O$_7$ ($R=$ Pr–Er) \cite{Li.arXiv.2501.14584}, corresponding to isovalent substitution of La$_3$Ni$_2$O$_7$. 
Interestingly, Sr-doped La$_3$Ni$_2$O$_7$ crystals were prepared under high pressure of 20 GPa at 1400$^{\circ}$C \cite{XuAdvElectMater2024}. 
Although the mechanism of crystal growth is unknown, application of high pressure seems reasonable considering that high Ni valence is stabilized under high pressures. 
The dimensions of the crystal were $0.059 \times 0.047 \times 0.031$ mm$^3$.

\subsection{Estimation of Oxygen Content}
The physical properties of La$_3$Ni$_2$O$_7$ and La$_4$Ni$_3$O$_{10}$ are sensitive to their oxygen contents, as mentioned later. 
The oxygen contents need to be determined with a precision on the order of 1/100. 
Since the formula weights of these compounds are approximately 646 and 892, respectively, the sample weight used for the determination must be measured with a precision of roughly 0.02\%, which is comparable to the repeatability of typical electronic balances even for relatively large sample masses of about 1 g.

Among common methods to estimate oxygen content or Ni valence, thermogravimetric (TG) analysis by hydrogen reduction or redox titration can generally achieve the required precision. 
Concerning TG analysis, however, conventional TG systems may not reach this accuracy because it is difficult to suppress fluctuations and thermal and time-dependent drifts. 
Such fluctuations and drifts may occur even if they appear to be removed by software provided with commercial systems. 
Therefore, weighing for oxygen-content estimation may need to be performed at room temperature to minimize errors due to air convection and change in buoyancy, using a high-repeatability balance with a large amount of pure sample, and in an inert atmosphere to prevent water absorption by La$_2$O$_3$ in the reduced sample \cite{SakakibaraPRB2024}. 
Water absorption, if it occurs, causes underestimation of oxygen content, which could explain why many reported values are lower than stoichiometric values. 
On the other hand, redox titration methods such as iodometry also seem unsuitable for these Ni oxides because oxygen bubbles are immediately generated when the sample is soaked in acid, indicating side reactions unrelated to the redox indicator.

Despite these challenges, reported oxygen contents are summarized here: $6.92$--$7.15$ for La$_3$Ni$_2$O$_7$ \cite{ZhangJSSC1994n1,AmowSSI2006RFR14,YuCEJ2016,LingJSSC1999,PoltavetsMRB2006,WuPRB2001,BannikovJSSC2006,WengJSSC2008,GaoApplMaterInterfaces2024,ChenPRL2024,WangPRX2024,SakakibaraPRB2024,UekiJPSJ2025} and $9.51$--$10.26$ for La$_4$Ni$_3$O$_{10}$ \cite{CarvalhoJMC1997,AmowSSI2006RFR14,YuCEJ2016,LingJSSC1999,ZhangJSSC1995,WuPRB2001,BannikovJSSC2006,WengJSSC2008,CarvalhoJSSC2009,SakakibaraPRB2024,NagataJPSJ2024}. 
Note that most reports do not describe the care taken in the determination, although values are typically reported with 1/100-order precision. 
In some cases, the samples used included secondary phases, meaning that the reported oxygen contents reflect only the average Ni valence of the sample. 
Even for single-phase samples, stacking faults may cause a distribution of Ni valence within the phase. 
The oxygen contents obtained by careful measurements are shown in Tables \ref{OxygenContent327} and \ref{OxygenContent4310}. 
As seen there, the oxygen contents of La$_3$Ni$_2$O$_7$ and La$_4$Ni$_3$O$_{10}$ after annealing at $p_{\mathrm{O_2}} = 1$ atm are nearly stoichiometric.

\begin{sidewaystable}
\caption{\label{OxygenContent327} Oxygen contents and Ni valences of La$_3$Ni$_2$O$_7$ annealed under various atmospheres.
}

\begin{tabular}{c|c|c|c|l}
Oxygen content & Ni valence & Atmosphere & Temperature & Note \\ \hline
7.01 & 2.51 & flowing O$_2$ & Slowly cooled from 1200$^{\circ}$C &  \\
7.00 & 2.51 & air & Slowly cooled from 600$^{\circ}$C &  \\
7.12 & 2.62 & $p_{\mathrm{O_2}}=300$ atm (HIP) & 600$^{\circ}$C & phase separation \\
7.17 & 2.67 & $p_{\mathrm{O_2}}=400$ atm (HIP) & 1200$^{\circ}$C & phase separation and decomposition \\
6.96 & 2.46 & flowing H$_2$ & 200$^{\circ}$C &  \\
6.84 & 2.34 & flowing H$_2$ & 250$^{\circ}$C & possible phase separation \\
6.50 & 2.00 & flowing H$_2$ & 300$^{\circ}$C &  \\
\end{tabular}

\caption{\label{OxygenContent4310} Oxygen contents and Ni valences of La$_4$Ni$_3$O$_{10}$ annealed under various atmospheres.
}
\begin{tabular}{c|c|c|c|l}
Oxygen content & Ni valence & Atmosphere & Temperature & Note \\ \hline
9.98 & 2.66 & flowing O$_2$ & Slowly cooled from 800$^{\circ}$C & without the La$_3$Ni$_2$O$_7$ intergrowth \\
9.94 & 2.63 & flowing Ar & Slowly cooled from 800$^{\circ}$C & without the La$_3$Ni$_2$O$_7$ intergrowth \\
9.99 & 2.66 & $p_{\mathrm{O_2}}=300$ atm (HIP) & 600$^{\circ}$C &  \\
10.04 & 2.69 & $p_{\mathrm{O_2}}=400$ atm (HIP) & 1200$^{\circ}$C & \\
9.88 & 2.59 & flowing H$_2$ & 200$^{\circ}$C &  \\
9.58 & 2.34 & flowing H$_2$ & 250$^{\circ}$C & multiphases \\
8.95 & 1.97 & flowing H$_2$ & 300$^{\circ}$C &  \\
\end{tabular}

\end{sidewaystable}

\subsection{Excess Oxygen}
\begin{figure}
\includegraphics[width=8cm]{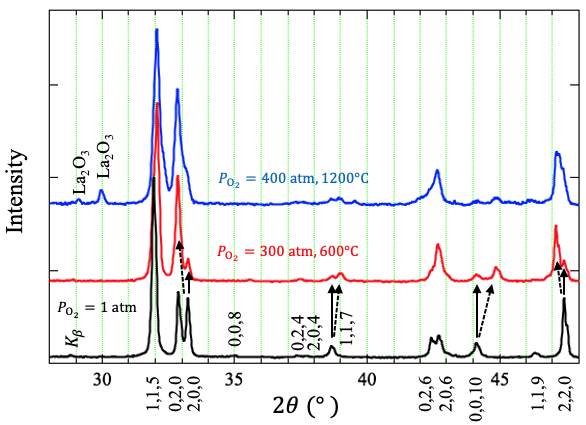}
\caption{\label{327XRD} 
XRD patterns of La$_3$Ni$_2$O$_{7.01}$, La$_3$Ni$_2$O$_{7.12}$, and La$_3$Ni$_2$O$_{7.17}$. 
Indexes are given assuming orthorhombic $Amam$ symmetry.
}
\end{figure}

Excess oxygen atoms can be introduced into the rock-salt slabs by annealing at high oxygen partial pressures, as shown in Tables \ref{OxygenContent327} and \ref{OxygenContent4310}. 
Interestingly, the introduction of excess oxygen into La$_3$Ni$_2$O$_7$ causes phase separation \cite{SakuraiSSP2025}. 
XRD patterns of La$_3$Ni$_2$O$_{7.01}$, La$_3$Ni$_2$O$_{7.12}$, and La$_3$Ni$_2$O$_{7.17}$ are shown in Fig. \ref{327XRD}, clearly indicating the phase separation of the latter two samples. 
For example, the 020 and 200 peaks of La$_3$Ni$_2$O$_{7.01}$ partially merge into single peaks in the other samples at $2\theta \simeq 32.85^{\circ}$, and the 0,0,10 peak at $2\theta \simeq 44.12^{\circ}$ for La$_3$Ni$_2$O$_{7.01}$ is partially shifted to $2\theta \simeq 44.88^{\circ}$, suggesting the coexistence of orthorhombic and tetragonal phases of La$_3$Ni$_2$O$_7$. 
This indicates that the oxygen content cannot be adjusted continuously and homogeneously above 7.01 around room temperature, similar to La$_2$NiO$_4$ as mentioned above. 
No phase separation or decomposition upon excess oxygen introduction was observed for La$_4$Ni$_3$O$_{10}$.

The weight ratio of the orthorhombic and tetragonal phases of La$_3$Ni$_2$O$_7$ was estimated to be approximately 35\% and 65\%, respectively \cite{SakuraiSSP2025}. 
Thus, assuming the composition of the orthorhombic phase is La$_3$Ni$_2$O$_{7.01}$ based on the nearly unchanged peak positions, the composition of the tetragonal phase can be calculated to be La$_3$Ni$_2$O$_{7.17}$ (= (LaO$_{1+1/6}$)(LaNiO$_3$)$_2$). 
This estimation seems reasonable because the excess oxygen of $\delta = 0.17$ exactly matches one of the $\delta$ values that cause commensurate superstructures in La$_2$NiO$_{4+\delta}$ (=(LaO$_{1+\delta}$)(LaNiO$_3$)) as mentioned in the previous section. 
The occurrence of such superstructures may drive the phase separation, although no superstructure has yet been observed for La$_3$Ni$_2$O$_7$. 
Interestingly, La$_3$Ni$_2$O$_{7.17}$ crystallizes in a tetragonal structure with higher symmetry than the original orthorhombic structure, whereas La$_2$NiO$_{4.25}$ exhibits reduced symmetry, from orthorhombic $Bmeb$ in La$_2$NiO$_4$ to monoclinic $C2$.

\subsection{Oxygen Deficiency}
Oxygen deficiency can be introduced mainly at the oxygen sites between two Ni sites along the $c$ axis (the inner apical oxygen sites), as mentioned in the previous section. 
However, the amount of deficiency cannot be widely controlled simply by varying $p_{\mathrm{O_2}}$ from 0 to 1 atm, as seen in Tables \ref{OxygenContent327} and \ref{OxygenContent4310}. 
This is in sharp contrast with high-$T_{\mathrm{c}}$ cuprates \cite{ETMJJAP1987}, although Ni and Cu ions in these oxides have valences higher than 2, the most common valence. 
To achieve significant reduction of the oxygen content in La$_3$Ni$_2$O$_7$ and La$_4$Ni$_3$O$_{10}$, more active reduction methods are required, such as hydrogen gas or CaH$_2$ treatment. 
The former is generally used to control oxygen content in La$_3$Ni$_2$O$_7$ and La$_4$Ni$_3$O$_{10}$, whereas the latter was used to synthesize La$_3$Ni$_2$O$_6$ from La$_3$Ni$_2$O$_7$ \cite{PoltavetsJACS2006}.

Oxygen atoms in La$_3$Ni$_2$O$_7$ and La$_4$Ni$_3$O$_{10}$ are indeed partially removed by heating in flowing hydrogen gas at appropriate temperatures. 
However, the amount of removed oxygen is not well reproducible at a given temperature. 
For example, the oxygen contents of La$_3$Ni$_2$O$_7$ samples were 6.35 and 6.84 when reduced at 450$^{\circ}$C in 11\% H$_2$/Ar and 11\% H$_2$/N$_2$, respectively \cite{ZhangJSSC1994n1}, strongly suggesting that the samples did not reach equilibrium. 
This is further supported by the fact that oxygen content depends on the duration of reduction: reduction in 10\% H$_2$/Ar at 450$^{\circ}$C yields La$_3$Ni$_2$O$_{6.38}$ after 12 hrs, but decomposition into La$_2$O$_3$ and Ni metal occurs after 60 hrs \cite{PoltavetsMRB2006}. 
The oxygen contents reported in Tables \ref{OxygenContent327} and \ref{OxygenContent4310} were obtained using samples reduced in approximately 5\% H$_2$/Ar for 12 hrs.

Temperature-dependent weight changes in thermogravimetric (TG) analysis by hydrogen reduction often exhibit step-like structures \cite{ZhangJSSC1994n1,ZhangJSSC1995,ChenPRL2024,LacorreJSSC1992,WangPRX2024,GaoApplMaterInterfaces2024,PoltavetsMRB2006}, although the exact shapes of the curves vary significantly between reports. 
These differences are most likely due to excessive heating rates. 
Because of the heating rate, all characteristic temperatures, such as decomposition temperatures, tend to be overestimated. 
For example, TG data suggest decomposition of La$_3$Ni$_2$O$_7$ into La$_2$O$_3$ and Ni metal occurs above 600–900$^{\circ}$C \cite{ZhangJSSC1994n1,ChenPRL2024,WangPRX2024,GaoApplMaterInterfaces2024,PoltavetsMRB2006}, although it can occur even at 450$^{\circ}$C, as noted above \cite{PoltavetsMRB2006}.
La$_3$Ni$_2$O$_7$ shows a single weight-loss step over approximately 100$^{\circ}$C around 500$^{\circ}$C, corresponding to oxygen contents of 6.35 \cite{ZhangJSSC1994n1}, 6.38 \cite{PoltavetsMRB2006}, and 6.45 \cite{ChengPRL2012,GaoApplMaterInterfaces2024}. 
Considering the phase at this step is particularly robust against temperature changes, it is likely that the oxygen content at the step is exactly 6.5, corresponding to purely divalent Ni ions in La$_3$Ni$_2$O$_{6.5}$. 
By contrast, La$_4$Ni$_3$O$_{10}$ exhibits two steps corresponding to La$_4$Ni$_3$O$_9$ and La$_4$Ni$_3$O$_8$ \cite{LacorreJSSC1992}, although the step structure is sensitive to heating rate and hydrogen concentration.

La$_3$Ni$_2$O$_{6.84}$ and La$_4$Ni$_3$O$_{9.58}$ listed in Tables \ref{OxygenContent327} and \ref{OxygenContent4310} likely correspond to intermediate states below the step temperatures.
Some XRD peaks of La$_3$Ni$_2$O$_{6.84}$ are broader than those of orthorhombic La$_3$Ni$_2$O$_{6.96}$ and tetragonal La$_3$Ni$_2$O$_{6.50}$, suggesting phase separation. 
For La$_4$Ni$_3$O$_{9.58}$, XRD patterns clearly show phase separation between the orthorhombic (or monoclinic) phase and tetragonal La$_4$Ni$_3$O$_9$. 
These observations indicate that oxygen deficiency in these compounds, as well as excess oxygen, is discontinuous in a single phase when prepared by simple reduction. 

\subsection{Aliovalent Substitution of La Ions -- Filling Control --}
Filling control is one of the most interesting and important challenges in superconductivity research. 
Aliovalent substitution of La ions, if feasible, would be particularly valuable, considering that filling control via oxygen content is strongly limited for the chemical reasons discussed above. 
Furthermore, random potentials introduced to the conducting electrons by aliovalent substitution of La ions are expected to be much smaller than those caused by oxygen defects; as mentioned above, excess oxygen atoms in the rock-salt slabs can tilt NiO$_6$ octahedra irregularly, and oxygen deficiency at the inner apical oxygen sites can reduce the hopping integrals between two Ni atoms along the $c$ axis.

Unfortunately, no tetravalent ions can substitute La ions, so there is no opportunity to reduce the Ni valence via La-site substitution. 
All tetravalent ions are too small for the La sites. 
By contrast, divalent alkaline-earth ions can partially substitute La ions in La$_3$Ni$_2$O$_7$. 
For A = Ca, Sr, and Ba, their contents $x$ in La$_{3-x}$A$_x$Ni$_2$O$_7$ have been reported to reach $x = 0.8$ \cite{ZhangJSSC1994n2,NedilkoJAC2004}, 0.2 \cite{ZhangJSSC1994n2,JiaoPhysC2024,XuAdvElectMater2024}, and 0.075 \cite{ZhangJSSC1994n2}, respectively. 
For Ca substitution, the electrical resistivity increases with increasing $x$ \cite{NedilkoJAC2004}, whereas Sr substitution reduces the resistivity at low pressures \cite{ZhangJSSC1994n2,JiaoPhysC2024}. 
For La$_{2.8}$Sr$_{0.2}$Ni$_2$O$_7$, however, the resistivity under high pressures becomes larger above approximately 10 GPa compared with La$_3$Ni$_2$O$_7$ \cite{XuAdvElectMater2024}. 
No superconductivity has been observed for aliovalent substitution of La ions in bulk samples.
For thin films, superconductivity has been observed in La$_{3-x}$Sr$_{x}$Ni$_2$O$_7$ ($0 \leq x \leq 0.21$) \cite{HaoNatMater2025}.

In general, careful examination is needed to ensure that a dopant is indeed incorporated into the target compound, especially when the dopant concentration is small and/or the sample contains secondary phases. 
Even if the dopant is incorporated, the dopant amount does not exactly correspond to carrier concentration, which also depends on the content of other elements, such as oxygen.

\section{Electronic Properties}
Studies on the physical properties of La$_3$Ni$_2$O$_7$ and La$_4$Ni$_3$O$_{10}$ have been vigorously pursued, despite experimental challenges in synthesizing high-quality samples and in conducting measurements under high pressure, particularly since the discovery of superconductivity in these compounds. 
Nevertheless, a complete consensus on their fundamental properties has not yet been reached, and research remains ongoing. 
Thus, hereafter, we briefly introduce representative experimental results and indicate directions for future studies. 

\subsection{Electronic Structure}
For La$_3$Ni$_2$O$_7$, first-principles calculations indicate that three types of electronic bands, denoted as the $\alpha$, $\beta$, and $\gamma$ bands, lie in the vicinity of the Fermi level \cite{PardoPRB2011,NakataPRB2017}.
This result is consistent with experimental observations obtained by synchrotron-based and laser angle-resolved photoemission spectroscopy (ARPES) \cite{YangNatComm2024}.
The $\alpha$ and $\beta$ Fermi surfaces have large Ni $3d_{x^2-y^2}$ orbital weight, although their bonding-antibonding splitting occurs due to the hybridization with the $3d_{z^2}$ orbitals. 
On the other hand, the $\gamma$ bands originate mostly from the bonding orbitals of Ni $3d_{z^2}$ orbitals. 

The $\alpha$ and $\beta$ bands form cylindrical Fermi surfaces extended along the $k_z$ direction, reflecting two-dimensional nature. 
The estimated doping levels are approximately 0.2 electrons/Ni and 1.25 holes/Ni, respectively \cite{YangNatComm2024}. 
According to band calculations without on-site Coulomb interactions \cite{PardoPRB2011,NakataPRB2017,YangNatComm2024}, these bands extend broadly from about $-1$ eV to 2.5 eV. 
Synchrotron ARPES revealed that they are narrowed by renormalization factors of about $\sim 2$ \cite{YangNatComm2024}.

\begin{figure}
\includegraphics[width=8cm]{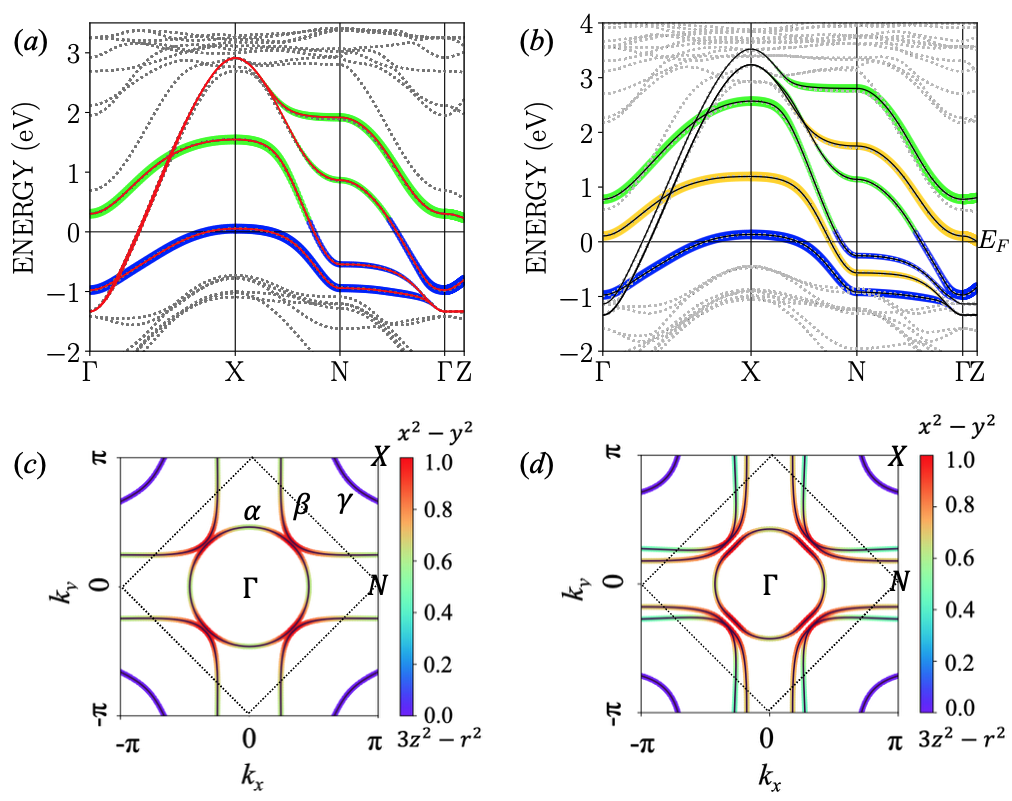}
\caption{\label{BandStruct} 
Electronic band structures of La$_3$Ni$_2$O$_7$ calculated using a four-orbital tight-binding model ($a$) \cite{NakataPRB2017}, and of La$_4$Ni$_3$O$_{10}$ using a six-orbital tight-binding model ($b$) \cite{SakakibaraPRB2024} 
Both calculations assume tetragonal $I4/mmm$ symmetry. 
Panels ($c$) and ($d$) show the Fermi surfaces at $k_z=0$ for La$_3$Ni$_2$O$_7$ and La$_4$Ni$_3$O$_{10}$, respectively. 
These panels were kindly provided by Prof. Sakakibara at Tottori University. 
The broken lines and the notations in panels ($c$) and ($d$) were added by the authors.
}
\end{figure}

The energy position of the $\gamma$ band relative to the Fermi level remains unsettled, and it is still under debate whether this band actually crosses the Fermi level to form a Fermi surface.
As shown in Fig.~\ref{BandStruct}, the top of the $\gamma$ band lies close to the Fermi level. 
Its energy is shifted downward by the inclusion of on-site Coulomb interactions, whereas it is shifted upward by the application of external pressure \cite{YangNatComm2024}.
These opposing effects make a definitive determination of the $\gamma$-band position difficult.
In addition, the $\gamma$ band is relatively flat and exhibits a strong tendency toward renormalization due to electron correlations.
The calculated bare bandwidth is approximately 1~eV, and the renormalization factor has been estimated to be in the range of 5–8 \cite{YangNatComm2024}, which is significantly larger than those for the $\alpha$ and $\beta$ bands.
Such large renormalization factors suggest that electron correlations are substantially stronger in the $\gamma$ band than in the $\alpha$ and $\beta$ bands.

The electronic band structure near the Fermi level can be well described by a four-orbital tight-binding model consisting of $d_{x^2-y^2}$-like and $d_{z^2}$-like Wannier orbitals derived from two Ni atoms stacked along the $c$ axis \cite{NakataPRB2017}, as illustrated in Fig.~\ref{BandStruct}.
In this model, tetragonal $I4/mmm$ symmetry is assumed, which leads to a notation of high-symmetry points in the Brillouin zone that differs from that of the orthorhombic $Amam$ structure.
In the $k_z$ plane, the orthorhombic Brillouin zone can be viewed as the tetragonal one folded along the lines connecting the midpoints of its edges (indicated by broken lines in Fig.~\ref{BandStruct}). 
This folding provides a geometric representation of the symmetry lowering from $I4/mmm$ to $Amam$.

Several features of the band structure can be qualitatively understood by considering a cluster consisting of two NiO$6$ octahedra stacked along the $c$ axis.
Because the NiO$_6$ octahedron is slightly elongated along the $c$ axis, the Ni $d{z^2}$ orbital lies lower in energy by $\Delta E$ than the $d{x^2-y^2}$ orbital.
For these two orbitals, a total of 1.5 electrons are expected to be distributed. 
Consequently, the $d_{z^2}$ orbital is half-filled, while the remaining 0.5 electrons occupy the $d_{x^2-y^2}$ orbital.
Furthermore, within the cluster, the two $d_{z^2}$ orbitals pointing toward the inner apical oxygen give rise to a large bonding–antibonding energy splitting through a sizable interlayer hopping integral, $t_{\perp}$.
As a result, the center of gravity of the $\gamma$ band is located at a lower energy than those of the $\alpha$ and $\beta$ bands and is (nearly) occupied, as shown in Fig.~\ref{BandStruct}.
This simple picture provides a reasonable description of the band structure along the $\Gamma$–$X$ direction, where hybridization between the $d_{z^2}$ and $d_{x^2-y^2}$ orbitals is forbidden by symmetry.
By contrast, along the $\Gamma$–$N$ direction, hybridization between these orbitals lifts the degeneracy of the $\alpha$ and $\beta$ bands, leading to the formation of distinct $\alpha$ and $\beta$ Fermi surfaces at the Fermi level.
Although this cluster-based picture is necessarily simplified, it highlights the crucial roles of $\Delta E$ and $t_{\perp}$ in shaping the low-energy electronic structure.

The electronic structure of La$_4$Ni$_3$O$_{10}$ \cite{SakakibaraPRB2024} can be understood within a similar cluster-based model.
Because its crystal structure is trilayered, three Ni atoms stacked along the $c$ axis must be considered.
Consequently, the coupling of three Ni $d{z^2}$ orbitals gives rise to a nonbonding orbital in addition to bonding and antibonding orbitals.
As a result, the electronic bands exhibit a slightly increased complexity, as illustrated in Fig.~\ref{BandStruct}.
Nevertheless, the overall electronic structure remains very similar to that of La$_3$Ni$_2$O$_7$.
Specifically, a relatively flat band is located around and just below the Fermi level, while more dispersive bands extend around and above it.

Based on these band structures, pronounced nesting features are expected \cite{NakataPRB2017,LuoPRL2023,LiNatComm2017,Du.arXiv2405.19853,Li.arXiv.2501.18885}, which is consistent with the observation of density-wave formations discussed in the following section.

\subsection{Density Wave}
\begin{figure}
\includegraphics[width=8cm]{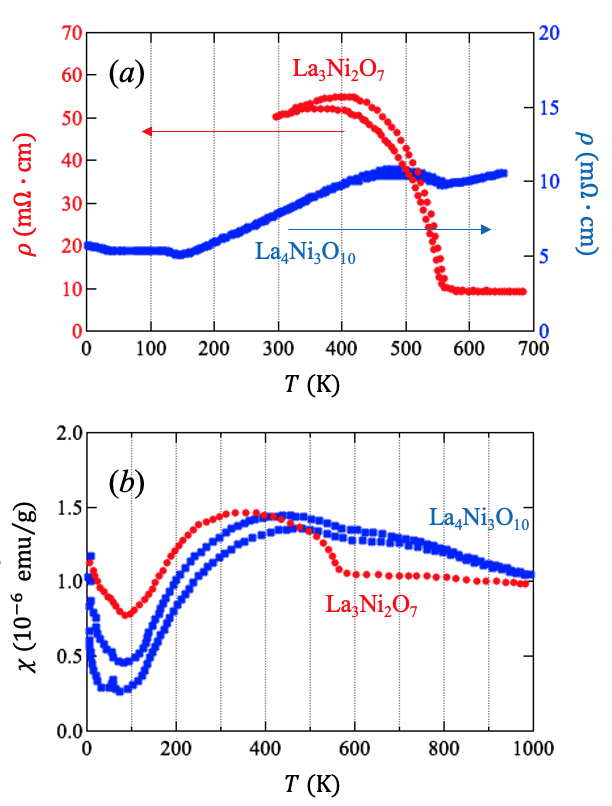}
\caption{\label{resist} 
Electronic resistivity ($a$) and magnetic susceptibility ($b$) of La$_3$Ni$_2$O$_7$ and La$_4$Ni$_3$O$_{10}$ as functions of temperature \cite{KobayashiJPSJ1996}. 
}
\end{figure}

At ambient pressure and zero field, La$_3$Ni$_2$O$_7$ exhibits metallic conductivity below approximately 700~K , as shown in Fig. \ref{resist}$a$ \cite{KobayashiJPSJ1996}. 
Around 550~K, a transition-like behavior from a good to a poor metallic state is observed, accompanied by changes in magnetic susceptibility (Fig. \ref{resist}$b$). 
The origin of this high-temperature transition remains unclear, although it may be related to the structural transition from the $I4/mmm$ to the $Amam$ space group reported near 700~K in some samples \cite{SongJMaterChemA2020}.

At lower temperatures, two additional transitions are observed at approximately 153~K and 110~K \cite{LiuSciChinaPhysMechAstro2022}. 
Below 153~K, the magnetic susceptibility perpendicular to the $c$-axis decreases slightly, accompanied by minor changes in resistivity. 
Below 110~K, the resistivity shows a modest increase without pronounced anisotropy in the magnetic response. 
Specific-heat anomalies are subtle due to the relatively high temperatures, but a small peak near 153~K is discernible. 
These low-temperature transitions are not universally observed in all samples, and the transition temperatures exhibit sample dependence. 
In some cases, the resistivity exhibits semiconducting behavior below 300~K or 140~K \cite{LingJSSC1999,CarvalhoJMC1997,ZhangJSSC1994n2,SreedharJSSC1994,ChenPRL2024,XieSciBull2024}, particularly in oxygen-deficient samples \cite{UekiJPSJ2025,JiaoPhysC2024,TaniguchiJPSJ1995}. 
Samples showing semiconducting behavior tend to exhibit upturns in the temperature dependence of the magnetic susceptibility at low temperatures \cite{LingJSSC1999,ZhangJSSC1994n2,WuPRB2001,XieSciBull2024}, likely due to oxygen vacancies that generate random potentials and localize magnetic moments on surrounding Ni sites. 
Consistently, the upturns were more pronounced in samples slightly reduced by hydrogen gas.

The 153~K and 110~K transitions are generally attributed to density wave formation. 
Although it was initially interpreted as charge density wave (CDW) formation \cite{WuPRB2001}, current evidence favors a spin density wave (SDW) scenario. 
NQR measurements on high-quality La$_3$Ni$_2$O$_7$ samples suggest a magnetic structure in which spins align ferromagnetically along the $a$-axis and antiferromagnetically along the $b$-axis within NiO$_2$ planes, while spins in adjacent planes along the $c$-axis are antiferromagnetically coupled \cite{YashimaJPSJ2025}. 
This model is almost consistent with muon spin rotation ($\mu$SR) \cite{ChenPRL2024}, resonant inelastic X-ray scattering \cite{ChenNatComm2024}, and neutron scattering experiments \cite{Plokhikh.arXiv.2503.05287}. 
Also in La$_2$PrNi$_2$O$_7$, magnetic order is observed below 161~K, as indicated by $\mu$SR and neutron diffraction \cite{Plokhikh.arXiv.2503.05287,KhasanovPRR2025}. 
In some samples, CDW–SDW coexistence or CDW has been suggested \cite{KakoiJPSJ2024,LuoChinPhysLett2025}, implying near-degenerate electronic states whose manifestation depends on crystallographic defects or sample-specific details.

Temperature-dependent measurements of the La nuclear spin-lattice relaxation rate divided by temperature (1/$T_1T$) exhibit divergent behavior near 150~K \cite{YashimaJPSJ2025}, consistent with a second-order density wave transition. 
Typically, both CDW and SDW transitions are of second-order with the gaps opening gradually below the transition temperatures as order parameters. 
However, optical conductivity studies indicate that the associated energy gap remains nearly constant below the transition \cite{LiuNatComm2024}, suggesting a sharp, first-order-like behavior. 
These discrepancies suggest that the low-temperature electronic state of La$_3$Ni$_2$O$_7$ is an open field for further investigation.

In La$_4$Ni$_3$O$_{10}$, metallic behavior persists below 700~K down to 150~K (Fig. \ref{resist}$a$) \cite{KobayashiJPSJ1996}. 
A small anomaly near 550~K may arise from minor inclusions of La$_3$Ni$_2$O$_7$ or stacking faults. 
Magnetic susceptibility gradually decreases above 400~K, resembling the behavior of La$_3$Ni$_2$O$_7$ between 300~K and 550~K. 
Considering that La$_4$Ni$_3$O$_{10}$ undergoes a structural transition to $I4/mmm$ at 873–1000~K \cite{SongJMaterChemA2020,NagellSSI2017} —approximately 200~K higher than La$_3$Ni$_2$O$_7$— a high-temperature metal–metal transition-like behavior similar to that in La$_3$Ni$_2$O$_7$ may occur above 700~K.

At low temperatures, La$_4$Ni$_3$O$_{10}$ also shows density wave formations.
Semiconducting behavior appears when large amount of oxygen deficiency exists as in the case of La$_3$Ni$_2$O$_7$ \cite{LiChiPhysLett2024} although the sample dependence is not so prominent.
For crystals with $P2_1/a$ symmetry, a transition occurs at 138.6~K, while $Bmeb$ crystals transition at 147.5~K \cite{ZhangPRM2020}. 
These transitions involve a temporary increase in resistivity below the transition temperatures \cite{LingJSSC1999,CarvalhoJMC1997,ZhangJSSC1995,ZhangPRM2020,ChenJACS2024,WuPRB2001,LiChiPhysLett2024,HuangfuPRR2020,LiCrystGrowthDesign2024,SamarakoonPRX2023} with anomalies in lattice-related properties such as lattice constants and thermal expansion coefficients \cite{ZhangPRM2020,KumarJMMM2020,RoutPRB2020}, suggesting CDW state. 
On the other hand, magnetic measurements reveal anisotropic responses to applied fields \cite{ZhangPRM2020,TakegamiPRB2024}, suggesting SDW formations.
In fact, complex density wave state of both spin and charge degrees of freedom has been proposed \cite{SamarakoonPRX2023,ZhangNatComm2020,Khasanov.arXiv.2503.04400v1}. 
Angle-resolved photoemission spectroscopy indicates a temperature-dependent gap, being consistent with second-order transition of the density wave formation \cite{Du.arXiv2405.19853}.
No experimental results show a first-order-like signatures. 
Substituting La with Pr or Nd does not significantly shift the transition temperatures \cite{ZhangJSSC1995,ZhangPRM2020,HuangfuPRR2020,RoutPRB2020,ZhangPRX2025n2}.

\subsection{Phase diagram}
The electronic states are determined by thermodynamic variables such as temperature ($T$), pressure ($P$), and magnetic field ($H$), and by the band filling. 
The fillings in La$_3$Ni$_2$O$_7$ and La$_4$Ni$_3$O$_{10}$ are directly related to oxygen stoichiometry. 
Thus, the electronic states of La$_3$Ni$_2$O$_7$ and La$_4$Ni$_3$O$_{10}$ can be considered functions of these four variables. 
In the following, we describe the electronic states of these compounds under representative combinations of these parameters. 
In many experimental studies, not all four variables are precisely controlled; in particular, oxygen content is often imperfectly defined. 
Consequently, some of the phase diagrams discussed here should be interpreted as projections of a four-dimensional electronic phase space.

\subsubsection{$P$--$T$ phase diagram}
The evolution of electronic and structural properties of La$_3$Ni$_2$O$_7$ and La$_4$Ni$_3$O$_{10}$ under pressure has been a central subject here. 
Of particular interest is how the structural transition from $I4/mmm$ to $Amam$ or $P2_1/a$, as well as the change from a good to a poor metal observed at ambient conditions, evolves at elevated pressures. 
Unfortunately, however, available $P$--$T$ phase diagrams extend only below 300~K, leaving the high-temperature regime unexplored. 

\begin{figure}
\includegraphics[width=8cm]{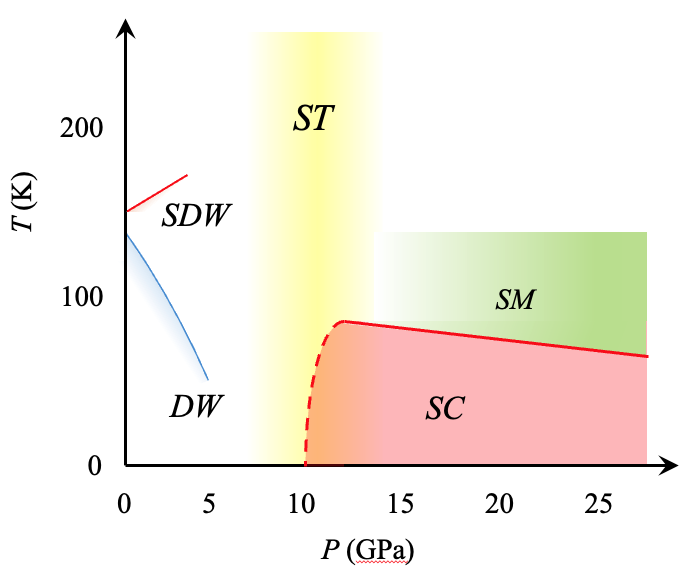}
\caption{\label{phasediagram} 
Schematic $P$--$T$ phase diagram of La$_3$Ni$_2$O$_7$.
$SDW$, $DW$, $ST$, $SM$, and $SC$ denote the spin-density wave, density wave, structural transition, strange metal, and superconducting phases, respectively.
}
\end{figure}

For La$_3$Ni$_2$O$_7$, as noted earlier, the sequential transitions at 153~K and 110~K have been observed at ambient pressure although they are sample-dependent and are not always clearly resolved. 
Nevertheless, a $P$--$T$ phase diagram such as that shown in Figure~\ref{phasediagram} has been broadly accepted \cite{SunNature2023,Li.arXiv2404.11369,ZhangNatPhys2024,ZhouMRE2025,WangPRX2024,ZhaoSciBull2025,ZhangJMaterSciTech2024,MengNatComm2024,KhasanovNatPhys2025}.
The higher-temperature transition of density wave formation shifts upward with increasing pressure, whereas the lower-temperature transition shifts downward. 
Both transitions vanish near the pressure at which the $Amam$ structure transforms into the $Fmmm$ phase. 
Although the structural phase boundary between $Amam$ and $Fmmm$ symmetries is not fully understood, it is generally assumed to be nearly vertical in the pressure axis \cite{WangJACS2024,WangInorgChem2025}. 
The first-order nature of the transition at the boundary implies the presence of a two-phase coexistence region, making the precise determination of phase boundaries challenging as in many cases. 
The $Amam$ structure becomes unstable above approximately 10~GPa, and so the $Fmmm$ phase likely starts to appear around 10~GPa. 
Although subtle superconductivity emerges in this regime, clear one appear above approximately 15~GPa, just above the structural transition pressure. 
The superconducting transition temperature, $T_{\mathrm{c}}$, is typically about 80~K there and decreases gradually with further compression. 
While density-wave states and superconductivity often compete in correlated systems such as organic conductors, in La$_3$Ni$_2$O$_7$ these two states are separated discontinuously by the first-order structural transition, leaving their relationship unresolved.

A strange metallic state has been reported just above $T_{\mathrm{c}}$, in which resistivity exhibits a linear temperature dependence \cite{SunNature2023,ZhangNatPhys2024,WangPRX2024,ZhangJMaterSciTech2024}. 
This has been interpreted as evidence for non-Fermi-liquid behavior and as a hallmark of unconventional superconductivity, as speculation from the high-$T_{\mathrm{c}}$ curate case. 
Although the superconductivity in La$_3$Ni$_2$O$_7$ is most likely unconventional indeed, key signatures of the strange metal observed in cuprates---such as abrupt changes in resistivity derivatives or Hall coefficients outside the strange metal regime---have not yet been demonstrated in La$_3$Ni$_2$O$_7$. 
Moreover, this strange metallic phase is not adjacent to a Mott insulating state as in the cuprates. 
Whether the observed metallic state is genuinely anomalous and whether it is directly related to the unconventional nature of superconductivity remain open questions. 
Some studies propose the presence of an additional density-wave phase in the strange-metal region \cite{MengNatComm2024}.

In a sample of La$_2$PrNi$_2$O$_7$, the structural transition sometimes occurs directly from $Amam$ to $I4/mmm$ near 11~GPa at room temperature \cite{WangNature2024}. 
This transition pressure is slightly lower than that of La$_3$Ni$_2$O$_7$, although the orthorhombicity increases owing to the samller ion of Pr as mentioned above.
Even more intriguing is the appearance of superconducting-like transitions at significantly lower pressures, around 8~GPa, well below the nominal structural transition.
While this may be attributed to local inhomogeneities that trigger structural changes at lower pressures, it raises the possibility that tetragonal symmetry may not be a strict prerequisite for superconductivity. 
As will be discussed below, similar behavior is reported in La$_4$Ni$_3$O$_{10}$ and Pr$_4$Ni$_3$O$_{10}$.

La$_4$Ni$_3$O$_{10}$ undergoes a pressure-induced structural transition from $P2_1/a$ to $I4/mmm$ around 15~GPa \cite{ZhuNature2024,ZhangPRX2025}.
A report claims that there is a $Bmeb$ phase between the $P2_1/a$ and $I4/mmm$ phase \cite{LiAdvMater2025}.
Beyond this transition, superconductivity emerges with a maximum $T_{\mathrm{c}}$ of 36~K \cite{NagataJPSJ2024}. 
Immediately after the discovery of the superconductivity in La$_4$Ni$_3$O$_{10}$ \cite{SakakibaraPRB2024}, the $P$--$T$ phase diagram was quickly constructed \cite{ZhuNature2024,ZhangPRX2025,LiAdvMater2025,LiChiPhysLett2024,XuNatComm2025}. 
Two kinds of $P$--$T$ phase diagram have been reported.
In the one, superconductivity appears below the structural transition pressure \cite{ZhuNature2024,ZhangPRX2025}, whereas in the other, it emerges just above the transition pressure \cite{LiAdvMater2025,LiChiPhysLett2024,XuNatComm2025}.
In either case, unlike La$_3$Ni$_2$O$_7$, the highest $T_{\mathrm{c}}$ is not found just above the transition pressure. 
The phase diagrams are more simple than that of La$_3$Ni$_2$O$_7$: there is a single transition of the density wave formation with transition temperature decreasing from around 140 K at ambient pressure with increasing pressure, although in a sample, magnetic structure change was observed at 90 K \cite{Khasanov.arXiv.2503.04400v1}.
Neither strange-metal behavior nor additional density wave phase at higher pressures has been observed in any $P$--$T$ region.

Pr$_4$Ni$_3$O$_{10}$ exhibits even more distinctive behavior \cite{ZhangPRX2025n2}. 
The structural transition to $I4/mmm$ occurs at approximately 35~GPa, yet superconductivity emerges already at around 25~GPa. 
Meanwhile, density-wave order persists up to pressures exceeding 40~GPa. 
Consequently, in the 25--40~GPa window, the system first undergoes a density-wave transition upon cooling, followed at lower temperatures by a superconducting transition. 
Although experimental uncertainties related to hydrostaticity at such high pressures necessitate further confirmation, this coexistence regime, if intrinsic, provides critical insights into the interplay between structure and competing electronic states in layered nickelates.

\subsubsection{$P$--$\delta$ phase diagram}
Precise determination of oxygen content in La$_3$Ni$_2$O$_7$ and La$_4$Ni$_3$O$_{10}$ remains scarce, yet several studies have examined the dependence of superconductivity on oxygen stoichiometry. 
In La$_3$Ni$_2$O$_7$, a reduction in oxygen content from the stoichiometric value of seven results in the immediate loss of metallic conductivity \cite{UekiJPSJ2025}. 
The system transforms into an insulating state, most likely driven by Anderson localization. 
As noted earlier, the missing oxygen resides within the perovskite block, and the resulting random potential strongly affects the NiO$_2$ planes. 
In contrast, excess oxygen exerts a different influence. 
When the oxygen content exceeds stoichiometry, the superconducting transition temperature $T_{\mathrm{c}}$ has been reported to remain as high as that of the stoichiometric sample below approximately 40~GPa, whereas it is markedly suppressed at higher pressures compared with the stoichiometric composition.
However, the interpretation is complicated by the fact that the powder samples used for these measurements were mixtures of orthorhombic La$_3$Ni$_2$O$_{7.01}$ and tetragonal La$_3$Ni$_2$O$_{7.17}$. 
At pressures below 40~GPa, the observed superconductivity may originate from La$_3$Ni$_2$O$_{7.01}$. 
Above 40~GPa, a distinct superconducting transition with lower $T_{\mathrm{c}}$ clearly emerges, which is associated with the oxygen-rich tetragonal phase. 
Supporting this interpretation, single crystals annealed under an oxygen partial pressure of 150 bar, yielding tetragonal La$_3$Ni$_2$O$_{7+\delta}$, did not exhibit superconductivity up to 68.2~GPa \cite{ShiarXiv.2501.14202}. 
These results suggest that oxygen excess shifts the superconducting onset to higher pressures and simultaneously reduces $T_{\mathrm{c}}$.

The situation is markedly different in La$_4$Ni$_3$O$_{10}$\cite{NagataJPSJ2024}.
For an oxygen-rich composition (La$_4$Ni$_3$O$_{10.04}$), superconductivity emerges at approximately 20~GPa, with sharp transitions observed beyond 30~GPa. 
The transition temperature increases to $T_{\mathrm{c}} \approx 36$~K at 48~GPa and then decreases gradually under further compression.
In contrast, in slightly oxygen-deficient La$_4$Ni$_3$O$_{9.99}$, superconductivity first appears near 30~GPa, but clear bulk transitions are only observed above 70~GPa. 
In this case, $T_{\mathrm{c}}$ reaches a maximum of 22~K at 79.2~GPa.
Thus, reducing the oxygen content shifts the onset of superconductivity to higher pressures and lowers the maximum $T_{\mathrm{c}}$, in contrast to the trend observed in La$_3$Ni$_2$O$_7$.
It should be noted, however, that the oxygen-deficient phase was synthesized by HIP annealing at 600$^{\circ}$C, which is lower than the temperature used to eliminate the La$_3$Ni$_2$O$_7$ stacking faults. 
This raises the possibility of residual stacking disorder, which could influence the superconducting properties.

\subsubsection{Other phase diagrams}
Combinations of $P$, $T$, $H$, and $\delta$ can, of course, give rise to phase diagrams beyond the $P$--$T$ or $P$--$\delta$ types.
Thus far, however, only $H$--$T$ phase diagrams have been reported, primarily to estimate the upper critical fields of superconductivity in La$_3$Ni$_2$O$_7$, La$_4$Ni$_3$O$_{10}$, and related compounds.
These results will be discussed in the final section.

\section{Thin Film}

Epitaxial thin films of La$_3$Ni$_2$O$_7$ have been fabricated by pulsed laser deposition (PLD) \cite{KoNature2025,Bhatt.arXiv.2501.08204,OsadaCommPhys2025,LiuNatMater2025}, molecular beam epitaxy (MBE) \cite{HaoNatMater2025,Zhong.arXiv.2502.03178}, and GOALL epitaxy methods \cite{ZhouNature2025}, providing an important platform to explore superconductivity under ambient pressure. 
When grown on substrates with smaller in-plane lattice constants than that of La$_3$Ni$_2$O$_7$, the films experience compressive strain within the $ab$ plane. 
For convenience, a pseudotetragonal lattice parameter of 3.833~\AA, defined as half the diagonal length of the $ab$ plane in the orthorhombic unit cell, is adopted as a reference value. 
On SLAO(001) [SrLaAlO$_4$(001)], whose lattice constant is approximately 2\% smaller than that of La$_3$Ni$_2$O$_7$, superconducting transitions with $T_{\mathrm{c}} = 26$--42~K at ambient pressure have been observed \cite{KoNature2025}. 
By contrast, films grown on NGO(001) [NdGdO$_3$(001)], which has a lattice constant about 0.7\% larger, remain metallic but nonsuperconducting, while those deposited on STO(001) [SrTiO$_3$(001)], with a lattice constant 1.9\% larger, exhibit semiconducting behavior \cite{Bhatt.arXiv.2501.08204}.

Beyond these three substrates, thin films have also been fabricated on LAST(001) [(LaAlO$_3$)$_{0.3}$(Sr$_2$TaAlO$_6$)$_{0.7}$(001)], with a relative lattice mismatch of $+0.9$\%, and on LAO(001) [LaAlO$_3$(001)], with $-1.2$\% \cite{KoNature2025,Bhatt.arXiv.2501.08204,Zhong.arXiv.2502.03178,OsadaCommPhys2025}. 
Films of substituted systems such as La$_{3-x}$Sr$_x$Ni$_2$O$_7$ ($x = 0$--0.45) and La$_{3-x}$Pr$_x$Ni$_2$O$_7$ ($x = 0.15, 1$) have also been synthesized \cite{HaoNatMater2025,ZhouNature2025,LiuNatMater2025}. 
Nevertheless, superconductivity at ambient pressure has been realized only in films grown on SLAO(001). 
It is worth noting that films grown on SLAO(100) often contain regions of La$_2$NiO$_4$, La$_4$Ni$_3$O$_{10}$, or amorphous-like domains, reflecting significant structural imperfections \cite{KoNature2025,Bhatt.arXiv.2501.08204}.

Several additional factors are crucial for achieving superconductivity in thin films.
One key parameter is film thickness: superconductivity has been observed only in ultrathin films, just several unit cells thick along the $c$ axis, most likely because sufficient substrate-induced strain is required.
Oxygen stoichiometry is another decisive factor. 
As-grown films exhibit nonmetallic behavior due to oxygen deficiency, and post-annealing in ozone is essential for restoring metallicity and superconductivity \cite{KoNature2025}.  
In La$_2$PrNi$_2$O$_7$ thin films, systematic annealing studies have shown that excessively high ozone concentrations decompose the films into perovskite phases, whereas insufficient annealing temperatures fail to induce superconductivity \cite{LiuNatMater2025}. 
Whether the absence of superconductivity after low-temperature annealing arises from excess oxygen incorporation or from incomplete oxygen uptake remains unresolved: at lower temperatures, ozone acts as a stronger oxidizing agent, while oxygen diffusion in oxides becomes slower.
Another important aspect is temporal stability \cite{KoNature2025}. Films that initially show zero resistivity after annealing tend to degrade over time, with resistivity above $T_{\mathrm{c}}$ gradually increasing, whereas re-annealing in ozone can restore superconductivity. 
This strongly suggests progressive oxygen loss.
Surface termination likely plays a role in stabilizing oxygen. 
In many superconducting films, a single unit cell of SrTiO$_3$ is deposited on the surface of La$_3$Ni$_2$O$_7$, which may help suppress oxygen escape.

The compressive strain imposed by SLAO(001) corresponds to an effective pressure of 10--20~GPa relative to bulk La$_3$Ni$_2$O$_7$. 
When plotted against the $a$-axis lattice constant, the superconducting transition temperatures of bulk and thin-film samples show a common trend: superconductivity emerges below $a \approx 3.79$~\AA\ and saturates at $T_{\mathrm{c}} \approx 60$--80~K for $a \lesssim 3.75$~\AA \cite{KoNature2025}. 
No such universal relation has been identified with respect to the $c$-axis parameter. 
Detailed structural analyses reveal that decreasing substrate lattice constants drive the La$_3$Ni$_2$O$_7$ layers from orthorhombic symmetry toward a more tetragonal structure, as expected from the compressive strain \cite{Bhatt.arXiv.2501.08204}. 
Importantly, even films grown on the other substrates, such as LAO(001), NGO(001), and STO(001) exhibit superconductivity once external pressure is applied, consistent with the $P$--$T$ phase diagram of bulk La$_3$Ni$_2$O$_7$ \cite{OsadaCommPhys2025}. 
These diagrams also display a density-wave transition at low pressures, with its transition temperature suppressed under pressure. 
Minor discrepancies remain, such as differences in the semiconducting regimes between bulk and thin films.
By contrast, the onset pressure for superconductivity does not systematically vary with the lattice mismatch among substrates: although SLAO(001) provides $-2$\% strain, NGO(001) $+0.6$\%, and STO(001) $+1.9$\%, the pressure required to induce superconductivity in each case remains similar \cite{OsadaCommPhys2025}. 
This observation suggests that substrate-induced strain alone cannot account for superconductivity in thin films. 
Other factors---such as oxygen stoichiometry or charge transfer across the film--substrate interface due to band alignment effects---may play equally significant roles.

\section{Superconducting properties and parameters}

At last, we would like to summarize the superconducting properties of La$_3$Ni$_2$O$_7$ and La$_4$Ni$_3$O$_{10}$. 
In bulk form, both compounds become superconducting only under very high pressures of several tens of GPa. 
Thus, the measurements to determine the superconducting parameters have been carried out using diamond anvil cells or multi-anvil cell, which imposes the measurements on various restrictions.
Consequently, precise measurements comparable to those at ambient pressure remain challenging. 
In addition, the high-quality sample well characterized is not always easy to obtain as mentioned above.
By contrast, in thin-film form, superconductivity has been observed at ambient pressure. 
However, such films are typically inhomogeneous, and the difficulty of determining sample mass in thin films prevents reliable evaluation of extensive quantities.
As a result, the superconducting properties of both La$_3$Ni$_2$O$_7$ and La$_4$Ni$_3$O$_{10}$ have been investigated primarily through electrical resistivity measurements. 
Magnetization measurements exist only in a limited number of cases, and no reports of specific heat measurements are available.

La$_3$Ni$_2$O$_7$ exhibits superconducting transitions even under extremely high magnetic fields, indicating that it is an extreme type-II superconductor. 
Penetration depth has been estimated for La$_{2.85}$Pr$_{0.15}$Ni$_2$O$_7$ film, to yield $\sim$4.5~$\mu$m at 1.8~K \cite{ZhouNature2025}, far exceeding the coherence lengths listed in Table \ref{SCpara}. 
The upper critical field $H_{c2}$ has been measured only up to 14~T, and higher values have been extrapolated using the empirical Ginzburg--Landau relation,
\[
  H_{c2}(T) = H_{c2}(0)\,\frac{1 - t^2}{1 + t^2}, \quad t = T/T_{\mathrm{c}} .
\]
Although the accuracy of such extrapolations is uncertain, it is generally accepted that $H_{c2}(0)$ is extremely large. 
To date, no direct evidence has been reported that $H_{c2}(0)$ exceeds the Pauli limit. 
Thin-film studies at ambient pressure show $T_{\mathrm{c}}$ values smaller than the maximum bulk values, yet coherence lengths of 1--2~nm, consistent with those inferred from bulk samples (Table \ref{SCpara}). 
Anisotropy of $H_{c2}$ has also been reported, though the difference between fields applied parallel and perpendicular to the $c$-axis is at most a factor of two. 
Substitution at the La site (e.g., by Pr or Sm) yields similarly high estimates of $H_{c2}$.

\begin{sidewaystable}
\centering
\caption{Superconducting parameters. $T_{\mathrm{c}}$, $H_{c2}(0)$, $H_{max}$, $P$, Criterion, $\xi$, and thickness represent superconducting transition at 0 T, upper critical field, the maximum magnetic field applied for the measurements, Pressure, the criterion of estimation of $T_{\mathrm{c}}$ related to the magnitude of the resistivity drop, coherence length, and thickness of the film estimated from the upper critical field, respectively. The coherence length is shown only when it is shown in the reference. (*: this value was estimated this time. In the paper, $\xi$ is estimated to be 4.83 nm.) }
\label{SCpara}
\begin{tabular}{|c|c|c|c|c|c|c|c|c|c|c|}
        \hline
        Composition & form & $T_{\mathrm{c}}$ (K) & $H_{c2}(0)$ (T) & $H_{max}$ (T) & $P$ (GPa) & Criterion (\%) & $\xi$ (nm) & Direction & thickness (nm) & Ref. \\ \hline
        La$_3$Ni$_2$O$_7$ & bulk & 80 & 158 & 7 & 20.1 & 90 & 	 & $c//H$ & 	 & \cite{ShiarXiv.2501.14202} \\ 
        	 & 	 & 72 & 73 & 	 & 22.6 & 	 & 	 & 	 &     & 	 \\ \cline{3-11}
        	 & 	 & 73 & 186 & 14 & 18.9 & 90  & 1.33* & $c//H$ & 	& \cite{SunNature2023} \\
        	 & 	 & 69 & 127 & 	 & 29.1 & 	 & 	 & 	 & & 	 \\
        	 & 	 & 65 & 97 & 	 & 43.5 & 	 & 	 & 	 & & 	 \\ \cline{3-11}
        	 & 	 & 67 & 86.6 & 8.5 & 14.5 & 90 & 	 &	 & 	 & \cite{WangPRX2024} \\ \cline{3-11}
        	 & 	 & 66 & 138 & 9 & 25.1 & 90 & 	 & 1.57 & 	 & \cite{ZhangNatPhys2024} \\
        	 & 	 & 71 & 97 & 	 & 20.5 & 	 & 1.84 & 	 & & 	 \\
        	 & 	 & 68 & 83 & 	 & 26.6 & 	 & 1.99 & 	 & & 	 \\ \hline
        La$_2$SmNi$_2$O$_7$ & bulk & 85 & 292 & 7 & 23.7 & 90 & 1.1 & 	 & 	 & \cite{Li.arXiv.2501.14584} \\ \hline
        La$_3$Ni$_2$O$_7$ & film & 38 & 132 & 14 & 0 & 90 & 1.8 & $c//H$ & 5 & \cite{KoNature2025} \\ 
        	 & 	 & 	 & 97 &  & 	 & 	 & 	 & $ab//H$ & & 	 \\ \hline        
        La$_{2.85}$Pr$_{0.15}$Ni$_2$O$_7$ & film & 34 & 119 & 14 & 0 & 90 & 2.2 & $c//H$ & 4.9 & \cite{ZhouNature2025} \\ 
        	 & 	 & 	 & 68 &  & 	 & 	 & 1.7 & $ab//H$ & & 	 \\ \hline  
        La$_2$PrNi$_2$O$_7$ & film & 48 & 142 & 14 & 0 & 90 & 1.7 & $c//H$ & 4 & \cite{LiuNatMater2025} \\ 
        	 & 	 & 	 & 116 &  & 	 & 	 & 	 & $ab//H$ & & 	 \\ \hline 
        La$_{2.91}$Sr$_{0.09}$Ni$_2$O$_7$ & film & 38 & 83.7 & 9 & 0 & 90 & 1.98 & $c//H$ & 5.21 & \cite{HaoNatMater2025} \\ 
        	 & 	 & 	 & 110.3 &  & 	 & 	 & 	 & $ab//H$ & & 	 \\ \hline
        La$_4$Ni$_3$O$_{10}$ & bulk & 18.5 & 24.4 & 34 & 50.2 & 50 & 3.3 & $c//H$ &   & \cite{Peng.arXiv.2502.14410} \\
        	 & 	 & 	 & 	 &  & 	 & 	 & 3.3 & $ab//H$ & & 	 \\  \cline{3-11}
        	 & 	 & 23 & 35 & 7 & 53 & 90 & 	 & $c//H$ & 	 & \cite{ZhuNature2024} \\
        	 & 	 & 25 & 44 & 	 & 63 & 	 & 	 & $c//H$ & 	 &  \\ \hline
        Pr$_4$Ni$_3$O$_{10}$ & bulk & 33 & 53 & 7 & 75.0 & 90 & 2.5 & $c//H$ &   & \cite{ZhangPRX2025n2} \\ \hline
\end{tabular}
\end{sidewaystable}

The highest reported onset-$T_{\mathrm{c}}$ for La$_3$Ni$_2$O$_7$ is 86~K \cite{ZhangJMaterSciTech2024}, while partial substitution yields up to 91~K in La$_2$SmNi$_2$O$_7$ \cite{Li.arXiv.2501.14584}, though the uncertainty in onset values is not small. 
Critical current densities $J_c$ have also been evaluated: bulk samples under 16.6~GPa show $J_c = 0.85$~kA/cm$^2$ at 1.5~K \cite{ZhangNatPhys2024}, while thin films yield $J_c = 0.32$~kA/cm$^2$ at 0.15~K (La$_3$Ni$_2$O$_7$) \cite{KoNature2025}, $J_c = 10.4$~kA/cm$^2$ at 1.4~K (La$_2$PrNi$_2$O$_7$) \cite{LiuNatMater2025}, and $J_c = 1.4$~kA/cm$^2$ at 2~K (La$_{2.91}$Sr$_{0.09}$Ni$_2$O$_7$) \cite{HaoNatMater2025}. 
These results suggest ample room for improvement, considering possible sample cracking under high pressure and the limited crystalline quality of films.

La$_4$Ni$_3$O$_{10}$ has also been considered to be an extreme type-II superconductor although neither lower critical field nor penetration depth have been measured.
Its upper critical field $H_{c2}$ has been determined directly over a wide range of applied fields \cite{Peng.arXiv.2502.14410}, yielding reliable results. 
The coherence length is estimated to be 3.3~nm. 
Anisotropy between in-plane and out-of-plane fields amounts to a factor of $\sim$1.5 down to 0.8$T_{\mathrm{c}}$, but diminishes at lower temperatures, converging to identical values at 2~K. 
This behavior provides an important information on assessing the reliability of extrapolated $H_{c2}$ values in La$_3$Ni$_2$O$_7$. 
Similar $H_{c2}(0)$ and coherence lengths have been reported in related compounds, including Pr$_4$Ni$_3$O$_{10}$ (Table \ref{SCpara}). 
The highest onset $T_{\mathrm{c}}$ values reported to date are 36~K for La$_4$Ni$_3$O$_{10}$ \cite{NagataJPSJ2024} and 40.5~K for Pr$_4$Ni$_3$O$_{10}$ \cite{ZhangPRX2025n2}.

Finally, we turn to the issue of superconducting symmetry in nickel oxides, a subject of broad interest across the research community.
Thus far, owing to the experimental challenges outlined above, no decisive data have been obtained for bulk samples.
In thin films of La-substituted La$_3$Ni$_2$O$_7$, ARPES and STM (scanning tunneling microscopy) measurements suggest nodeless gap structures \cite{Shen.arXiv.2502.17831,Sun.arXiv.2507.07409}, but the precise symmetry of the superconducting order parameter remains unresolved.
Clarifying this issue is essential not only for establishing the nature of the superconducting state, but also for uncovering the underlying pairing mechanism.

\section*{Acknowledgment}

We are grateful to our collaborators, in particular Prof. Kazuhiko Kuroki (University of Osaka), Prof. Hirofumi Sakakibara (Tottori University), Prof. Masayuki Ochi (University of Osaka), and Prof. Hidekazu Mukuda (University of Osaka), for their invaluable support and insightful discussions.
The SEM and EDX images shown in Fig.~\ref{La2O3andNi} were obtained by Mr. Kazuki Yamane, and the crystal structure illustrations in the figures were prepared using VESTA \cite{MommaJAC2011}.
This work was partly supported by the World Premier International Research Center Initiative (WPI), MEXT, Japan, and by JSPS KAKENHI (Grant Nos. JP24K01333 and JP25K00959).


\end{document}